\documentclass[11pt]{article}
\usepackage[margin=1in]{geometry}
\usepackage[hyphens]{url}      
\usepackage{graphicx}
\usepackage{natbib}
\usepackage{caption}
\usepackage{enumitem}
\usepackage{amsmath}
\usepackage{booktabs}
\usepackage{amsfonts}
\usepackage{algorithm}
\usepackage{algorithmic}
\usepackage{newfloat}
\usepackage{listings}
\usepackage{libertine}
\usepackage[hidelinks,breaklinks]{hyperref}
\hypersetup{
    colorlinks=true,
    linkcolor=cyan,
    filecolor=magenta,
    urlcolor=cyan,
    citecolor=cyan,
}

\title{An External Fairness Evaluation of LinkedIn Talent Search}
\author{
  Tina Behzad\textsuperscript{1},
  Siddartha Devic\textsuperscript{2},
  Vatsal Sharan\textsuperscript{2},\\\vspace{3ex}
  Aleksandra Korolova\textsuperscript{3},
  David Kempe\textsuperscript{2}\\
  \vspace{1.5ex}
  \textsuperscript{1}Stony Brook University \quad
  \textsuperscript{2}University of Southern California \quad
  \textsuperscript{3}Princeton University
}

\setcitestyle{round}
\newcommand\blfootnote[1]{
    \begingroup
    \renewcommand\thefootnote{}\footnote{#1}
    \addtocounter{footnote}{-1}
    \endgroup
}

\begin{document}

\maketitle

\begin{abstract}
We conduct an independent, third-party audit for bias of LinkedIn's Talent Search ranking system, focusing on potential ranking bias across two attributes: \emph{gender} and \emph{race}. 
To do so, we first construct a dataset of rankings produced by the system, collecting extensive Talent Search results across a diverse set of occupational queries. We then develop a robust labeling pipeline that infers the two demographic attributes of interest for the returned users.
To evaluate potential biases in the collected dataset of real-world rankings, we utilize two exposure disparity metrics: deviation from group proportions and $\text{MinSkew}@k$. 
Our analysis reveals an under-representation of minority groups in early ranks across many queries.
We further examine potential causes of this disparity, and discuss why they may be difficult or, in some cases, impossible to fully eliminate among the early ranks of queries. 

Beyond static metrics, we also investigate the concept of subgroup fairness over time, highlighting \emph{temporal disparities} in exposure and retention, which are often more difficult to audit for in practice.
In employer recruiting platforms such as LinkedIn Talent Search, the persistence of a particular candidate over multiple days in the ranking can directly impact the probability that the given candidate is selected for opportunities. Our analysis reveals demographic disparities in this temporal stability, with some groups experiencing greater volatility in their ranked positions than others.
We contextualize all our findings alongside LinkedIn’s published self-audits of its Talent Search system and reflect on the methodological constraints of a black-box external evaluation, including limited observability and noisy demographic inference. 
Our work contributes empirical insights and practical guidance for conducting third-party audits of modern socio-technical systems which go beyond the well-studied and standard algorithmic fairness guarantees of predictors.
\end{abstract}

\blfootnote{
Contact: tbehzad@cs.stonybrook.edu, \{devic, vsharan\}@usc.edu, korolova@princeton.edu, David.M.Kempe@gmail.com
}
\blfootnote{This work appeared at AAAI 2026.}

\newpage
\tableofcontents

\newpage

\section{Introduction}

\begin{figure}
    \centering
    \includegraphics[width=1\columnwidth]{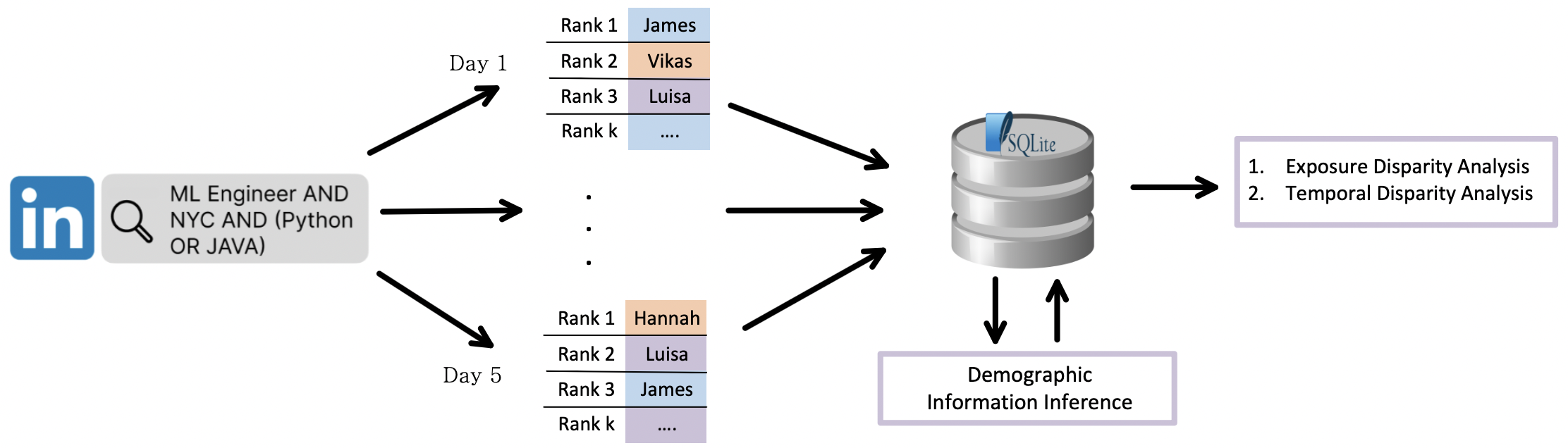}
    \caption{A schematic overview of our pipeline: we issue identical queries to LinkedIn Talent Search over five consecutive days (Section \ref{sec:data_retrieval}), ingest the results into our database, and then enrich these records with demographic inferences using external APIs and datasets (Section \ref{sec:data_labeling}). Finally, we carry out exposure‐disparity analysis (Section \ref{sec:exposure_disparity_analysis}) and temporal‐disparity analysis (Section \ref{sec:temporal_analysis}).}
    \label{fig:enter-label}
    \vspace{-5pt}
\end{figure}
LinkedIn is one of the most important platforms for hiring around the world. According to LinkedIn’s official statistics, more than 10,000 members worldwide apply for jobs on the platform every minute \citep{linkedin_about_statistics}.
LinkedIn's employer-focused recruiting suite, LinkedIn Recruiter, is one of the platform's most impactful aspects, serving as an influential and widely-used tool through which employers find potential job candidates.
LinkedIn's Recruiter platform provides a host of features that connect employer recruiters to relevant candidates for particular roles, via a robust candidate search and screening filters.
According to LinkedIn, more than 5.7 million talent professionals across 1.1 million companies use LinkedIn Recruiter to source and hire candidates, and seven people are hired through the platform every minute \citep{linkedin_talent_solutions}.

LinkedIn operates at a massive scale, and has a tangible impact on the modern labor market.
Given this, the\emph{fairness} of its recruiting platform remains as important an issue as ever. 
Previous work has demonstrated that some components of social media platforms, such as Meta's employment \citep{ali2019discrimination, imana2021auditing} and education \citep{imana2024auditing} ad delivery mechanisms, can be biased.
However, to the best of our knowledge, the key aspect of LinkedIn's recruiting platform responsible for directly connecting job market candidates to recruiters, its \emph{recruiter-facing candidate search tool}, has not yet been independently studied.

\begin{figure*}
    \centering
    \includegraphics[scale=0.4]{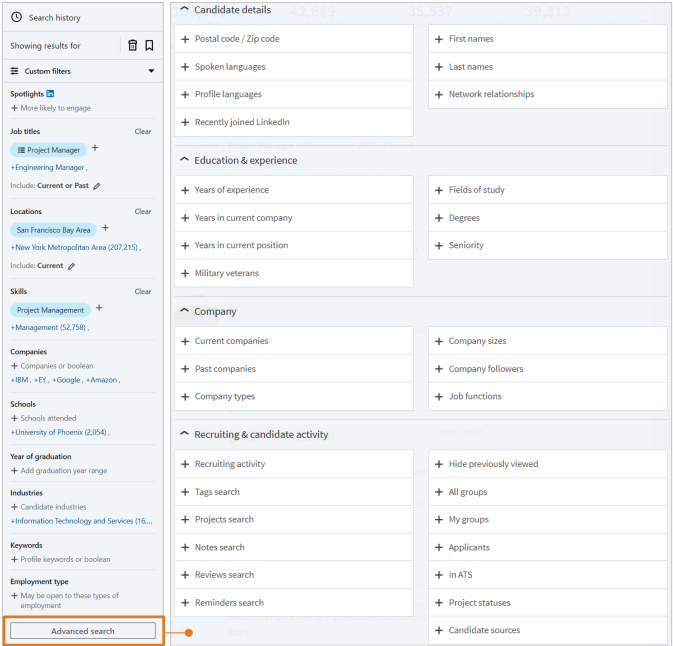}
    \caption{Candidate search filters available in LinkedIn Recruiter. Source: \citet{linkedin_search_filters}.}
    \label{fig:recruiter-search-filters}
\end{figure*}

LinkedIn Talent Search (LTS) allows recruiters to view a \emph{ranking} over individuals when searching for candidates using a particular \emph{query}, which consists of a combination of skills and other criteria specified by the recruiter.
For example, a recruiter for a technology company in New York City may construct an LTS query that searches for nearby candidates with skills in Python and at least three years of experience in management. Figure~\ref{fig:recruiter-search-filters} presents a short representation of possible search filters. 

Given that recruiters have limited resources and place considerable trust in LinkedIn's algorithms,\footnote{Independent surveys report LinkedIn as the leading social recruiting channel, with over 71\% of 1,200 respondents using it \citep{nextthingrpo2024employ}.} the resulting ranking over candidates for any particular LTS query thus has considerable power in shaping who gets seen, contacted, and ultimately hired. 
Given LinkedIn's dominance, LTS can influence both individual careers and the diversity of the broader corporate workforce.
Minute unfairness or even unintended behaviors on the order of a single-digit percentage can have lasting downstream impacts on thousands of people.

Although LinkedIn's self-reports claim improvements across defined fairness metrics \citep{geyik2019fairness}, independent and replicable external audits are crucial for providing validation of these efforts and improving trustworthiness \citep{longpre2025house}.
Our goal is to conduct an independent, replicable external audit of LinkedIn Recruiter’s ranking algorithms for potential biases across demographic groups. In addition, we will advance the methodological toolkit \citep{metaxa2021auditing} needed for overcoming practical barriers to carry out such audits effectively \citep{Cen-transparency, casper2024black, imana2023having}.

\subsection{Overview of Audit}
We conduct an external, fully independent empirical audit of LinkedIn's Recruiter platform candidate search system, LTS, and compare our findings against publicly available LinkedIn's self-reports of its fairness initiatives.
First, in Section~\ref{sec:LinkedIn-ranking}, we elaborate on the challenges of collecting data in the absence of internal or full platform access to LinkedIn Recruiter Search. 
Next, in Section~\ref{sec:data_retrieval}, we address the task of data collection and labeling, focusing in particular on the difficulty of inferring demographic information from candidate profiles. 
In Sections~\ref{sec:exposure_disparity_analysis} and \ref{sec:temporal_analysis}, we discuss and select fairness metrics that capture meaningful notions of equity across multiple time scales and for multiple subgroups, often in the face of incomplete or noisy data.
At a high level, our results point to disparities between candidates with differing genders in (1) the early ranks of queries (the first 100 ranks / the first 4 pages of candidate results); and (2) when repeating queries over multiple days.
Our work serves both as an analysis of fairness outcomes in LTS and as a reflection on the broader challenges of performing such external audits.

\section{Related Work}
\subsection{Fairness in Ranking Algorithms}
A growing body of research addresses fairness in ranking and recommendation systems, focusing on metrics and algorithms to ensure that ranked results do not systematically disadvantage certain groups \citep{patro2022fair, rastogi2024fairness, mehrotra2021mitigating, mehrotra2022fair, devic2024stability}. 
Early works define fairness on a single ranking instance, for example, by requiring each top-$k$ prefix of a ranked list to have an adequate representation of protected groups \citep{singh2018fairness}. \citet{zehlike2017fa} propose one such constraint that each prefix of the ranking contains at least a minimum fraction of items from the disadvantaged group. 

Beyond early prefix-based definitions, subsequent work has introduced more nuanced formulations of fairness in ranking that account for exposure, relevance, and stability across multiple instances.
\citet{singh2018fairness} propose a framework to ensure that, over many ranking instances, items from each group receive exposure proportional to their merit or relevance.
\citet{liu2018personalizing} develop fairness-aware re-ranking methods that balance exposure between groups while maintaining recommendation quality, demonstrating the practical trade-off between fairness and utility.
\citet{rastogi2024fairness} define a new fairness criterion that corresponds to a group-wise fair lottery among relevant options, even in the presence of disparate uncertainty.
Finally, \citet{devic2024stability} highlight stability as a complementary fairness dimension, showing that even when exposure is balanced, large fluctuations in rankings across iterations can disproportionately affect disadvantaged groups.

\citet{joseph2016fairness} explore how a natural fairness principle, such as ``similar individuals should be treated similarly'', can be upheld in sequential decision-making. They design algorithms that delay favoring one option over another until there is strong enough evidence to justify doing so, ensuring that fairness is respected even when the system is still learning. They prove that their method satisfies formal fairness constraints while still achieving sublinear regret. 
\citet{wang2021fairness} introduce a different fairness objective focused on the equitable allocation of exposure among items in stochastic bandits. They propose a method that incorporates merit-based fairness of exposure into stochastic bandits and demonstrate that their approach achieves both fairness and utility guarantees. Additionally, \citet{mehrotra2021mitigating} account for noise in inferring protected attributes by formulating a “denoised” optimization problem and offering a linear programming–based solution that improves fairness metrics despite the presence of uncertain attribute labels.

While much of the fairness literature assumes a fixed snapshot of data and outcomes, deployed systems rarely operate in such static conditions. Real-world models interact with users and data streams, producing feedback loops that shape future distributions. Understanding how fairness interventions behave and persist over time is therefore essential.

Several foundational works in fair machine learning have begun to model such temporal dynamics explicitly. In classification settings, \citet{liu2018delayed} introduce a one-step feedback model showing that optimizing solely for immediate accuracy can lead to long-term disparities. In the labor market domain, \citet{hu2018short} demonstrate that short-term fairness constraints can yield fair and stable long-term outcomes, underscoring the impact of early-stage interventions on system-level equity.

More recent studies examine how fairness evolves in dynamic ranking and recommendation systems. \citet{cao2024recommendation} study fairness evolution in social recommendation networks and find that visibility disparity tends to decrease over time, suggesting that recommendation fairness can improve as systems evolve. \citet{davis2025emerging} analyze fairness drift, showing that predictive models may become less equitable as they evolve, even when fairness constraints are initially satisfied. \citet{10.1145/3514094.3534173} examine the long-term consequences of fairness interventions in connection recommendation systems. Using simulations and theoretical analysis, they show that common exposure and utility parity interventions can inadvertently amplify bias over time, producing a group-wise rich-get-richer effect. Although such interventions appear fair in aggregate, they fail to promote equity in network sizes, leaving minority groups increasingly disadvantaged in the long run. Together, these studies emphasize that fairness is not static but evolves dynamically, motivating the need to evaluate and maintain fairness longitudinally in ranking and recommendation systems.

Despite these advances, \citet{patro2022fair} argue that most fairness efforts still focus on one-shot decisions, with limited attention to cumulative or long-term effects as rankings and predictions are repeatedly applied. Even with the growing body of work highlighting the downstream and temporal consequences of unfair decisions, to the best of our knowledge, no metric has been specifically designed to quantify the unfairness of rankings themselves over time. This gap is particularly important, as repeated exposure to lower ranks can compound disadvantages for certain groups \citep{patro2022fair}. Developing such a metric represents a critical step toward identifying and mitigating cumulative harms in ranking-based systems.

\subsection{LinkedIn’s Fairness Research in Candidate Ranking}
LinkedIn has acknowledged the importance of ensuring fair and equitable candidate rankings and has invested significantly in improving its algorithms over the years. 
For instance, the work by \citet{geyik2019fairness} presents LinkedIn's early efforts in quantifying and mitigating algorithmic bias, focusing on gender and age disparities in candidate rankings in particular. 
Based on LinkedIn's self-report, such efforts significantly improved the visibility and representation of historically underrepresented candidates without negatively impacting key business metrics.
More recently, LinkedIn has published work outlining their holistic approach to operationalizing fairness at scale, focusing on system-level practices rather than isolated algorithms \citep{Hsu2023disentangling}. LinkedIn frames its AI fairness agenda as the combination of equal AI treatment and product equity. The equal AI treatment component includes auditing and evaluating fairness through metrics such as predictive parity \citep{10.1145/3593013.3594117}, conducting root-cause analyses, exploring mitigation strategies that do not rely on demographic information, and, when the use of demographic information is justified, employing it cautiously to address disparities while monitoring for unintended consequences. On the product equity side, they highlight interventions such as diversity nudges, which notify recruiters when a group is underrepresented in search results and suggest modifications to search filters to improve representation.\footnote{We were not able to observe this diversity nudge in any of our searches.}

\subsection{External Audits for Fairness and Accountability}
While a growing body of research on fair ranking has introduced a variety of metrics and algorithmic interventions (e.g., \citet{zehlike2017fa, geyik2019fairness, singh2018fairness}), most of these methods are evaluated solely on benchmark datasets such as academic search corpora, synthetic data, or static logs from recommender systems. Such evaluations often fail to capture the complexities and dynamics of real-world deployment, including feedback loops, user adaptation, and institutional context. This gap between simulation and deployment raises critical questions about whether fairness interventions generalize beyond curated test settings. It is therefore essential to study deployed systems, particularly those that have publicly committed to fairness goals, to assess whether these interventions produce meaningful change at scale. While platforms are often well-positioned to conduct fairness audits internally and in some cases already do so \citep{geyik2019fairness}, \citet{longpre2025house} argue that independent, third-party audits remain vital for accountability. Internal evaluations may downplay or overlook harms, whereas external audits have repeatedly surfaced discrepancies between public fairness claims and actual outcomes.
Several prominent studies demonstrate the impact of external audits on big online platforms. For example, independent researchers showed that Facebook’s ad delivery algorithm was skewed along demographic lines, even when advertisers did not target by those attributes. In experiments, identical job ads were delivered disproportionately to certain groups, reflecting biases by gender, race, and age in how the platform’s algorithm learned to distribute ads \citep{imana2021auditing}. In another work, \citet{wang2024lower} conduct a randomized field study comparing Twitter/X’s algorithmic timeline to its chronological counterpart. By analyzing over 800,000 tweets shown to 243 users over multiple weeks, they audit the platform’s real-world content delivery and user experience. Their findings reveal that the algorithmic feed surfaces fewer news items overall, but with slightly higher reliability and less ideological extremity compared to the chronological timeline. These cases show how external audits can rigorously evaluate real-world system behavior, uncover latent biases, and provide meaningful evidence about the consequences of algorithmic design choices.

A key challenge in conducting external audits is the need to infer demographic information to assign group memberships before any fairness metric can be meaningfully evaluated. Some external audits have leveraged publicly available U.S. voter registration lists, which include demographic attributes \citep{ali2019discrimination,imana2021auditing}. Another common approach involves inferring demographic information from indirect signals. For instance, Bayesian Improved Surname Geocoding (BISG) is widely used to estimate race based on a combination of surname and geographic location \citep{elliott2009using,badrinarayanan2024privacy}. We discuss this in more detail in Section~\ref{sec:data_labeling}.

\section{LinkedIn's Ranking Pipeline} \label{sec:LinkedIn-ranking}

LinkedIn Recruiter (LTS) operates using a two-stage ranking architecture, a standard approach in large-scale information retrieval systems \citep{liu2009learning, dang2013two}.
\footnote{Our description of LinkedIn’s talent search is based on the latest publicly available academic and technical publications. We acknowledge that some of these sources are several years old and that the actual systems or algorithms used by LinkedIn may have evolved since their publication.}
There is a large pool of candidates $C$. For any particular query $q$, the platform's goal is to present the recruiter with a ranked list of relevant candidates. Given the scale of LinkedIn’s user base, computing a full ranking over all possible candidates in $C$ is prohibitively expensive.

In the first stage, known as candidate retrieval, a small subset of relevant candidates $C_q \subseteq C$ is selected from the overall pool $C$. The set $C_q$ represents candidates that may be of interest to a recruiter issuing query $q$.
This retrieval stage is powered by \emph{Galene} \citep{LinkedInGalene}, LinkedIn’s in-house search engine.
Galene generates an initial list of candidates based on a feature-level matching utilizing candidate profile aspects such as job titles, skills, employment history, etc.

In the second stage, LinkedIn uses machine-learned models to assign a relevance score to each candidate in $C_q$. 
A common paradigm within the second stage is for a score function $r^*(x)$ to map each individual $x \in C_q$ to a \emph{relevance score} in $[0,1]$; the individuals are then returned to the recruiter in order of decreasing score.
One important property of LinkedIn's relevance score is that it captures a combination of two aspects: (1) LinkedIn's estimate that the individual $x$ is qualified for $q$; \emph{and} (2) LinkedIn's estimate that the candidate is willing to respond to the query $q$.
LinkedIn's motivation for incorporating both components into the relevance score is to satisfy the recruiters: neither unqualified nor uninterested/unmovable candidates would be useful to recruiters, and a large percentage of such candidates would lead to recruiters' dissatisfaction with LTS.\footnote{According to LinkedIn, the ``InMail'' acceptance rate of candidates accepting such connections from interested recruiters is a key business metric for the company \citep{10.1145/3269206.3272030}.}

The ranked lists should not only be useful, but also adhere to hiring laws and best practices.\footnote{We mainly focus on US laws and hiring practices in this work, since the queries we make are conducted from and geographically centered within the US.}
Operationalizing hiring laws and best practices is often ambiguous given the uncertainty about how interviewing and hiring laws apply to digital hiring pipelines.\footnote{See, e.g., the ongoing lawsuit Derek Mobley v.~Workday Inc.~(Case 3:23-cv-00770).}
Nonetheless, LinkedIn has released information \citep{geyik2019fairness} about the fairness metric(s) they aim to optimize and some of the interventions that they apply before presenting the final ranked list to the recruiter.
We describe one fairness metric and the corresponding post-processing method (\textsc{DetGreedy}) which, as of 2019, LinkedIn applies to all returned candidate rankings within recruiter search.

Suppose that the overall candidate pool $C$ can be partitioned into $m$ \emph{disjoint} groups $C = g_1 \cup g_2 \cup \dots \cup g_m$ based on sensitive attributes or profile data.
Once the relevant candidate list $C_q$ is obtained, the proportion of each group $g_i$ in query $q$ is defined as $p^*_i = | C_q \cap g_i | / |C_q|$.
Notice that $p^*_i$ represents the proportion of the \emph{returned relevant candidates} which belong to group $g_i$, not the proportion of candidates from the entire pool $C$.
Let $c_i^k$ denote the cumulative count of the number of individuals belonging to group $g_i$ from ranks 1 through $k$ (inclusive) in the ranking shown to the recruiter.
\citet{geyik2019fairness} propose a suite of \emph{post-processing} algorithms which, for every group $g_i$, at each $k \in \{ 1, \dots, |C_q|\}$, enforce:
\begin{align*}
    \lfloor p^*_i \cdot k \rfloor \leq c_i^k \leq \lceil p^*_i \cdot k \rceil.
\end{align*}
After defining and experimenting with several proposed algorithms in \citet{geyik2019fairness}, LinkedIn chose to deploy \textsc{DetGreedy} in LTS. This algorithm has theoretical guarantees of satisfiability 
when the number $m$ of groups is at most 3.
Since the post-processing algorithm is applied \emph{after} the individuals are sorted by score, individuals are therefore never compared between different groups $g_i \neq g_j$; the scores are only used to rank the individuals \emph{within} their own group. For each position in the ranking, the candidate with the highest score from the currently most underrepresented group is selected.
The fact that such an approach works relies crucially on the disjointness of groups; it would be a much more difficult task if the groups were overlapping, as is desired in other fairness notions for candidate ranking \citep{dwork2019learning,devic2024stability}.

\section{Data Collection}
\subsection{Retrieving Ranking Data} \label{sec:data_retrieval}
\begin{figure*}
    \centering
    \includegraphics[scale=0.25]{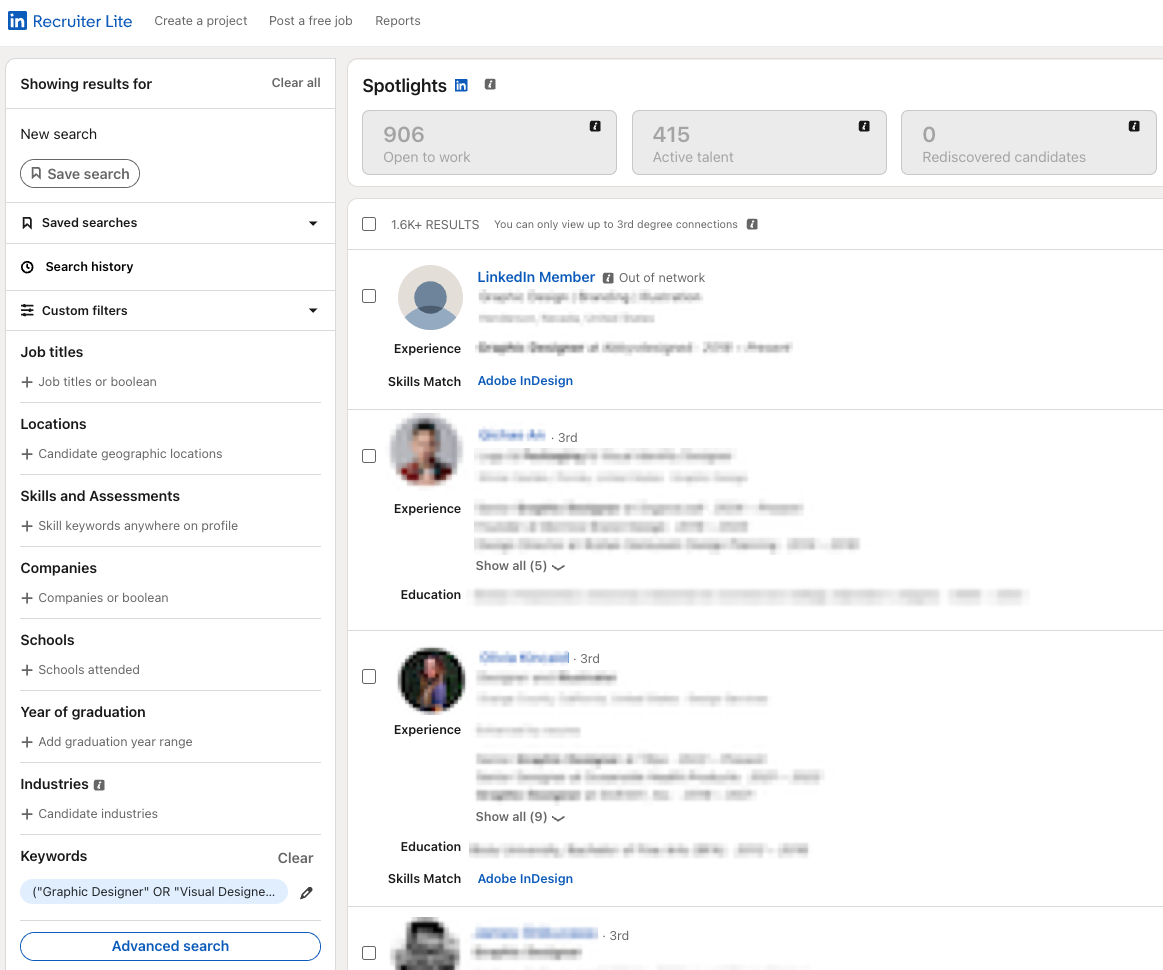}
    \caption{A snapshot of the results for a sample query where the top-ranked candidate is missing. Candidate cards are blurred to preserve privacy.}
    \label{fig:linkedin_member}
\end{figure*}

In this section, we describe how we collected data from LinkedIn's \emph{Recruiter Lite} platform.\footnote{\url{https://business.linkedin.com/talent-solutions/recruiter-lite}}
We note that Recruiter Lite has more limited functionality and access than the full Recruiter platform.
However, obtaining access to the full Recruiter platform requires: (1) being a well-established company; (2) a meeting with a LinkedIn representative; and (3) paying an undisclosed amount on the order of many thousands of dollars per recruiter per year --- requirements that are virtually impossible to satisfy for independent auditors.
In contrast, the Recruiter Lite platform can be accessed with a free trial and can be extended beyond that for \$170 a month.
Recruiter Lite, to the best of our knowledge and experience, can be easily requested by any real individual with an active and populated LinkedIn profile and a plausible story.
As we performed the audit without LinkedIn's express permission or knowledge, and did not have access to any company's full Recruiter platform license, we performed our investigations with Recruiter Lite.
Recruiter Lite memberships are subject to several important limitations:
\begin{itemize}
\item \textbf{Limited search scope}: Search results are restricted to the account owner's 3rd-degree connections. Assuming that each LinkedIn user has around 200 connections,\footnote{LinkedIn claims that the average number of connections per person in the US was 109 in 2016 \citep{barbarasa2017skills}.} with an estimated 20\% overlap in 2nd-degree connections and 30\% overlap in 3rd-degree connections, this results in a reachable network of approximately 4.5 million people.

\item \textbf{Daily candidate limit}: The number of unique candidates viewable in LTS queries is capped at 2,000 per day. This means that collecting a substantial dataset requires multiple days of queries.

\item \textbf{Result page limit}: Only the first 1,000 candidates (40 pages of 25 candidates each)
are viewable for any query.

\item \textbf{Hidden candidate data}: Candidates outside the account holder’s 3rd-degree connections appear in search results as \textit{LinkedIn Member} with anonymized details (see Figure~\ref{fig:linkedin_member} for an example). Since these candidates are ``missing'' data, we skip them when scraping. Even if their inclusion were to offset disparities, recruiters using Recruiter Lite would not be able to view or contact these anonymized candidates, so they would not contribute to balancing representation in practice.

\item \textbf{No access to job-seeking status}: Recruiter Lite membership does not provide visibility into whether a candidate is actively looking for a job by marking themselves as ``open to work''. This limitation prevents us from tracking transitions between active and passive job-seeking states. These signals could serve as proxies for successful hires or help explain why certain candidates disappear from search results over time.
\end{itemize}

We employ the Selenium Python package and its built-in Chrome WebDriver to automate data collections (see below discussion for additional details). We note that all data collected are publicly available on the LinkedIn Recruiter platform.
We collected rankings of around 26,000 candidates across 78 different queries. All queries were collected from the state of New York without a VPN.\footnote{The Recruiter product could potentially behave differently in different countries; our audit is focused on the US version.}
We also perform many of the same queries across multiple days to investigate the \emph{temporal} aspect of the rankings. 

We categorize queries into three levels of search specificity: general queries, general queries with full candidate card information, and position-specific queries with full candidate card information.

\paragraph{Query Set $Q_1$: General Queries.} These queries were selected from the job titles listed by the U.S. Bureau of Labor Statistics (BLS),\footnote{https://www.bls.gov/cps/cpsaat11.htm}

ensuring that group-level statistics were available for each. In total, we selected 40 titles, aiming to include at least one representative from each major category or subcategory defined by the BLS to preserve diversity across professions. Each query was repeated over five consecutive days. Using these titles as keyword queries, we applied the “NYC Metropolitan Area” location filter and scraped ranked candidates up to rank 200 (eight pages). For each query, we collected all data visible on the candidate cards, including profile name, headline, connection level, profile picture link if available, and background details (education, experience, and skills) without clicking on the "see more" option. We were able to scrape data for up to only eight queries per day. These queries were conducted from December 30th, 2024 to January 28th, 2025. In the following sections we refer to this query set as $Q_1$.

\paragraph{Query Set $Q_2$: General Queries with Full Candidate Card Information.} From February 12th to March 28th, 2025, we executed a set of 40 queries. For each query, we collected results up to the 200th rank on five consecutive days within this time period. During this phase, we clicked each candidate card’s “See more” button to extract the full list of their experience, background and skills. We refer to this query set as $Q_2$.  While $Q_2$ contains the same set of queries as $Q_1$, the key difference is that $Q_2$ includes full candidate card information. In the remainder of this paper, we focus our analysis on $Q_2$ for general queries.

\paragraph{Query Sets $Q_3$ and $Q_3^{\prime}$: Position Specific Queries with Full Candidate Card Information.} To mirror real-world Recruiter searches, we collected descriptions of random LinkedIn job postings (fetched within an Incognito window without logging in to LinkedIn) and prompted OpenAI's o4-mini-high model to generate a query from the posting’s requirements. 
Each query consists of four to five AND-connected terms, where each term itself OR-combines several related skills or job titles.
The resulting query returns a focused set of qualified candidates for that job posting. 
We then iteratively refined each query manually until it returned fewer than $1,000$ qualified candidates, allowing us to retrieve the complete ranked list within the constraints of Recruiter Lite. 
For every candidate in the ranked lists, we collected the same information as in Set 2, $Q_2$. For 9 of these queries, data were retrieved on a single day; for the remaining 15, they were collected over five consecutive days. These queries were conducted from March 30th to April 2nd, 2025. 
In the following sections, we refer to this query set as $Q_3$. To minimize the impact of missing candidates in the rankings, we selected 8 queries for which the proportion of LinkedinIn members was less than 1\% of the total results. Identifying queries that met this criterion was particularly challenging, as LinkedIn frequently returns rankings that include inaccessible members. For each of these queries, we collected the same data for all returned candidates between May 13th and May 23rd, 2025. In the following sections we refer to this query set as $Q_3^{\prime}$.

\paragraph{Scraping Details.}
Our scraping approach involves the following two steps: (1) After logging in and opening Recruiter, the code automatically enters the given query into the search box, which by default performs a keyword search. 
(2) The code then scrolls through the results, scraping the personal information displayed for each candidate.\footnote{The code is available at \url{https://github.com/tina-behzad/LinkedIn-Audit}}
Because LinkedIn does use interactions with a profile to refine estimates of the searcher's preferences and adjust future rankings accordingly, our code avoids any interactions with profiles. 
That is, it restricts scraping to the information available on each immediately accessible \emph{candidate card} (see Figure~\ref{fig:linkedin_member} for example anonymized candidate cards). 
When a \emph{see more} link appears in a card’s Experience, Education, or Skills sections, we click it to expand and capture the additional information in the list. 
This was applied for all query batches below except the first.
Clicking ``see more'' is unlikely to influence the rankings since \emph{explicit} modes of recruiter interactions, such as clicking and spending time on the profile, contacting or saving the candidate for later, etc.,  are usually considered more important.

\subsection{Data Labeling} \label{sec:data_labeling}
Accurately assessing fairness in algorithmic systems typically requires member-level demographic signals, for example, gender, race, or ethnicity, in order to measure disparities across different groups. 
For external audits like ours, this means devising a method to infer demographic attributes without direct access to ground truth labels. 
LinkedIn faces a similar challenge internally as only about 6\% of LinkedIn's U.S. members have self-reported their race/ethnicity via LinkedIn’s Self-ID survey \citep{badrinarayanan2024privacy}. 
To expand fairness assessments beyond this limited group, LinkedIn has developed a Privacy-Preserving Probabilistic Race/Ethnicity Estimation (PPRE) method \citep{badrinarayanan2024privacy}. 
This system combines Bayesian Improved Surname Geocoding (BISG, see \citet{elliott2009using}), data from the Self-ID survey, and privacy-enhancing technologies, including secure two-party computation and differential privacy. 
Crucially, this approach avoids assigning deterministic race/ethnicity labels to individuals. Instead, it generates probabilistic estimates on the fly, which are encrypted, aggregated, and immediately discarded after measurement. This design ensures that fairness evaluations can be conducted responsibly, without compromising individual privacy or enabling downstream use of inferred demographic attributes.

In the following discussion, we outline the strategies we used to assign gender and race labels for candidates in our audit. 

\paragraph{Gender Estimation.} In our analysis, we focus on binary gender (male/female).\footnote{We acknowledge that this framing does not capture the full spectrum of gender identities, and we recognize the limitations this imposes on the inclusivity and comprehensiveness of our results.}
We use candidates' first name to infer their gender using the Social Security Administration's national name dataset \citep{ssa-babynames}.
For each name, we assign the gender with the higher frequency based on the \emph{Count} column in the dataset, which indicates how many times the name was recorded for each gender. 
When a name appears with only one associated gender, the assignment is straightforward. For names not found in the dataset, we leverage the free tiers of publicly available commercial APIs ( \citet{genderAPI} and \citet{genderizeIO}) to infer gender using the first name. Using the combination of these three sources, we labeled the vast majority of candidates, and only 600 individuals (2.3\%) remained unlabeled. 
To validate the accuracy of our gender labels, two of the authors manually annotated 5,700 candidate profiles using full names and profile pictures (when available).\footnote{These manual labels were not used in our analysis and were only used to measure the accuracy of our inference.} 

Across the full set of 5,700 manually labeled cases, the agreement with the automatically assigned labels was 95\%.

\paragraph{Race Estimation.}
As mentioned above, LinkedIn uses a variation of the BISG method to infer racial categories, combining full name and geolocation data. Since we do not have access to candidate geolocation information, this approach was not feasible. 
Instead, we restrict our inference to first and last names. We first applied the Nameparser Python package to clean and standardize names and titles, and then used the Ethnicolr Python package \citep{ethnicolr} 
to assign racial labels based on first and last name. The package classifies individuals into four racial categories: Asian, Hispanic, Non-Hispanic White, and Non-Hispanic Black based on census data and voter registration records. To validate its accuracy, we independently annotated 400 individuals using their full name and profile picture. The agreement between the two annotators was 82\%, and the average agreement between the annotators and the automated labels was 65.5\%. Given the limited reliability of the automated labels across all groups, we restricted our analysis to only Non-Hispanic White and an aggregated ``Other'' category. This increased the annotators' agreement to 88\% and the average agreement with automated labels to 72\%.

Although such annotator agreement is far from perfect, it is in line with that of auditing methods relying on inference for protected attributes \citep{elliott2009using} and consistent with expectations. Meta’s technical report on race inference using BISG methods reports a maximum accuracy of 85\%.\footnote{\url{https://ai.meta.com/research/publications/how-meta-is-working-to-assess-fairness-in-relation-to-race-in-the-us-across-its-products-and-systems/}}
Techniques for accounting for noise in inference when measuring machine-learning bias is an active area of research (e.g., \citet{mehrotra2022fair, imana2025auditing, rastogi2024fairness, ghosh2023fair}).

\section{Analysis} \label{sec:exposure_disparity_analysis}
In this section, we evaluate the effectiveness of potential post-processing methods in LTS using two key metrics that measure disparities in group representation. We focus on gender analysis, with results for racial groups included in Appendix~\ref{app:results_gender}. Since both metrics require accurate estimates of each group's overall proportion, we limit our analysis to position-specific queries where the full candidate pool is available (query set $Q_3$ and $Q_3^{\prime}$), allowing for more reliable estimation.
In Appendix~\ref{app:results_gender}, we provide the same analysis for the general query set $Q_2$.

\subsection{Deviation from Group Proportion} 
\label{subsec:group-deviation}

In \citet{geyik2019fairness}, LinkedIn noted that they apply the \textsc{DetGreedy} algorithm (as described in Section~\ref{sec:LinkedIn-ranking}) using binary gender as the grouping variable. Under this approach, the cumulative count for each group $i$ at rank $k$ should approximately match $p^*_i \cdot k$, where $p^*_i$ represents the share of group $i$ in the overall retrieved candidate pool. 
In other words, the proportion of candidates from each group should closely reflect $p^*_i$ throughout the ranked list.

For the position-specific queries, $Q_3$ and $Q_3^\prime$, we can compute observed group proportions directly from the data, and use them as an estimate of $p_i^*$ for evaluating exposure disparities by calculating deviation from the true proportion.

The total number of retrieved candidates for queries in $Q_3$ ranges from 73 to 1,000, the upper limit under our Recruiter Lite membership. We restrict the analysis to queries with less than 15\% missing candidates.\footnote{The missing candidate rate thresholds for each set were chosen to balance data quality with maintaining adequate query coverage. Recall that candidates are ``missing'' in the Recruiter Lite rankings if they are further than three hops in the connection graph from the recruiter (Section~\ref{sec:data_retrieval}).} Here, we examine ranks up to 300 for queries in $Q_3$. Throughout this analysis and the remainder of Section \ref{sec:exposure_disparity_analysis}, we rely on the candidate rankings retrieved on the first day each query was run. While these queries provide access to what appears to be the full pool of returned candidates, the presence of even a small number of missing profiles can introduce noise into the estimation of the true group proportions $p_i^*$. To improve accuracy, we looked for queries with less than 1\% missing data and scraped the complete candidate lists for each. We retrieved data for eight such queries ($Q_3^\prime$), with candidate pool sizes ranging from 111 to 698. Here, and in the subsequent plots, we examine ranks up to 200 for queries in $Q_3^\prime$ since six of the eight queries contain fewer than 200 candidates. The full query text and corresponding candidate pool sizes for both sets are listed in Appendix \ref{app:position_specific_queries}.

\begin{figure}
    \centering
    \includegraphics[width=1\linewidth]{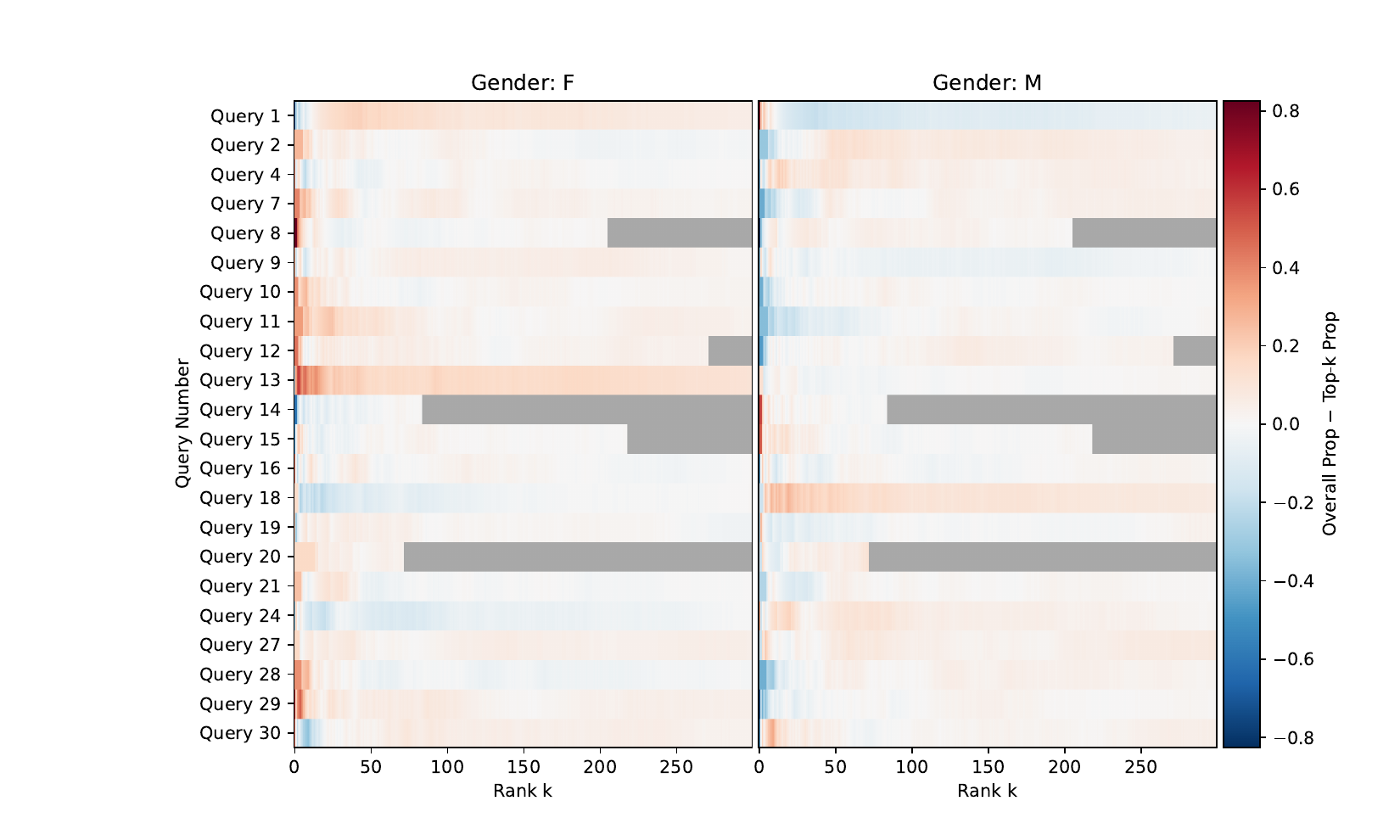}
    \caption{Deviation between the observed top-$k$ gender proportions and the overall candidate pool proportions for the set of queries for which we scraped the full list of returned candidates ($Q_3$). Each row corresponds to a query, with gender-wise deviations shown across rank positions up to $k = 300$. Gray areas indicate ranks beyond the total number of returned candidates for that query (i.e., the candidate pool was smaller than 300). Red values indicate under-representation relative to the overall group proportion, while blue values indicate over-representation.}
    \label{fig:specific_queries_gender_deviation}
\end{figure}

\begin{figure}
    \centering
    \includegraphics[width=1\linewidth]{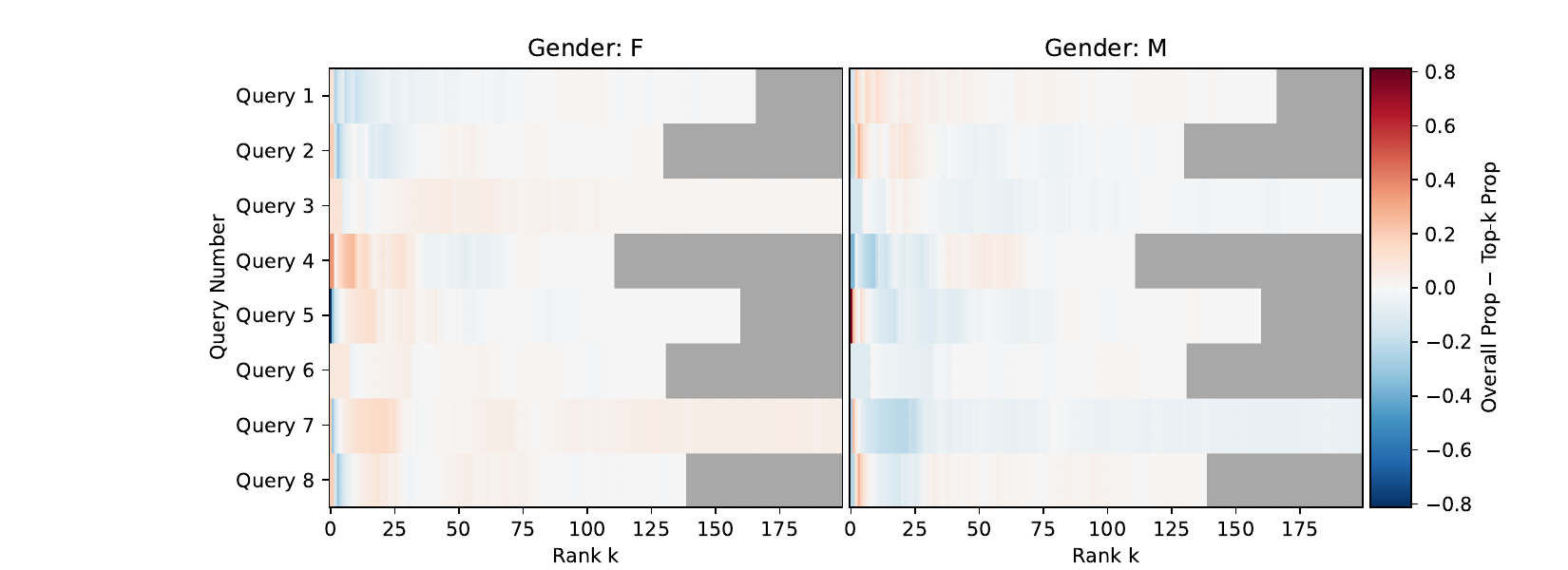}
    \caption{Deviation between the observed top-$k$ gender proportions and the overall candidate pool proportions for the set $Q'_3$ of queries with less than 1\% missing members. Gender-wise deviations are shown across rank positions up to $k = 200$. Gray areas indicate ranks beyond the total number of returned candidates for that query.}
    \label{fig:specific_queries_gender_deviation_no_missing}
\end{figure}

Figure \ref{fig:specific_queries_gender_deviation} and Figure \ref{fig:specific_queries_gender_deviation_no_missing} visualize the deviation $p^*_i - \tfrac{n_{i,k}}{k}$ for each query in $Q_3$ and $Q'_3$, respectively, where $n_{i,k}$ denotes the cumulative count of candidates from group $i$ up to rank $k$ and $p^*_i$ denotes the overall proportion of group $i$ in the candidate pool.

Both plots consistently show that the magnitude of deviation is greatest in the very top ranks and then rapidly diminishes toward zero further down the list.\footnote{We use the terms top ranks and early ranks interchangeably throughout the paper, both referring to candidates appearing at the beginning of the rankings.} At very low $k$, the deviation bars (deep red or blue) are largest, but by $k \approx 100$ they have largely collapsed toward white. In other words, small slices of the top of the ranking are where gender skew is most pronounced; as more candidates are being considered, the sample proportions gravitate back toward the overall baseline.
For the female panel (left), nearly every query shows a red bias at $k \leq 25$, indicating under-representation of women in the very top slots relative to their overall share. By contrast, the male panel (right) is mostly dominated by blue at the very top.

After about {$k \geq 75$}, almost all entries are near zero, so by the third or fourth page of results, there is essentially no gender skew. However, this does not diminish the impact of disparities observed at the top of the ranking. Algorithmic systems are subject to position and trust biases; users are more likely to engage with candidates near the top and to trust that those candidates are the most qualified \citep{10.1145/1229179.1229181, 10.1145/1076034.1076063}. In platforms like LinkedIn Recruiter, where recruiters typically begin reviewing from the top of the list and may never reach lower-ranked pages, skew at early ranks carries the most weight. Even small disparities at the top can translate into meaningful inequities in visibility and opportunity.

\subsection{Skew@$k$}\label{sec:result_skew}
Although computing deviation from the expected group proportions at top-$k$ ranks is a standard approach to measure statistical parity, and aligns with standard metrics such as Normalized Discounted Difference proposed by \citet{yang2017measuring}, in the work \citet{geyik2019fairness}, LinkedIn introduced an alternative metric known as \emph{MinSkew@k}.

This metric quantifies the worst-case deviation from the target representation among all protected groups at a given cutoff $k$. Specifically, the skew for group $g_i$ in the ranked list $\tau_r$ is defined as:
\begin{equation}
\text{Skew}_{g_i}@k (\tau_r) = \log \left( \frac{p^{\tau_r}_{k,g_i}}{p^{q}_{g_i}} \right),
\end{equation}
where $p^{\tau_r}_{k,g_i}$ denotes the proportion of candidates with attribute value $g_i$ in the top-$k$ of the ranked list $\tau_r$, and $p^{q}_{g_i}$ is the desired (target) proportion for $g_i$ in the given query $q$. Subsequently, MinSkew@k is defined as:
\begin{equation}
\text{MinSkew}@k(\tau_r) = \min_{g_i \in G} \text{Skew}_{g_i}@k(\tau_r),
\end{equation}
capturing the most disadvantaged group's deviation at rank $k$. 
According to the reported results of \citet{geyik2019fairness}, the average $\text{MinSkew}@100$ improves from $-0.259$ to $-0.011$ after applying the \textsc{DetGreedy} algorithm. They also report similar improvements across other cutoffs, such as top-25 and top-50, with $\text{MinSkew}@k$ values approaching zero consistently for over 95\% of the queries. 

To interpret the change, recall that $\text{MinSkew}@k$ is defined as the logarithm of the ratio between the observed and target group proportions at rank $k$. A value of $-0.259$ corresponds to a group being represented at roughly 77\% of its expected share (since $\exp(-0.259) \approx 0.77$), whereas $-0.011$ indicates near parity, with the observed share being approximately 99\% of the target ($\exp(-0.011) \approx 0.989$). See Appendix \ref{app:minskew} for a detailed, step-by-step example of how this metric is calculated and how it varies as the underlying distribution changes.

\begin{figure}[h]
    \centering
    \includegraphics[scale=0.5]{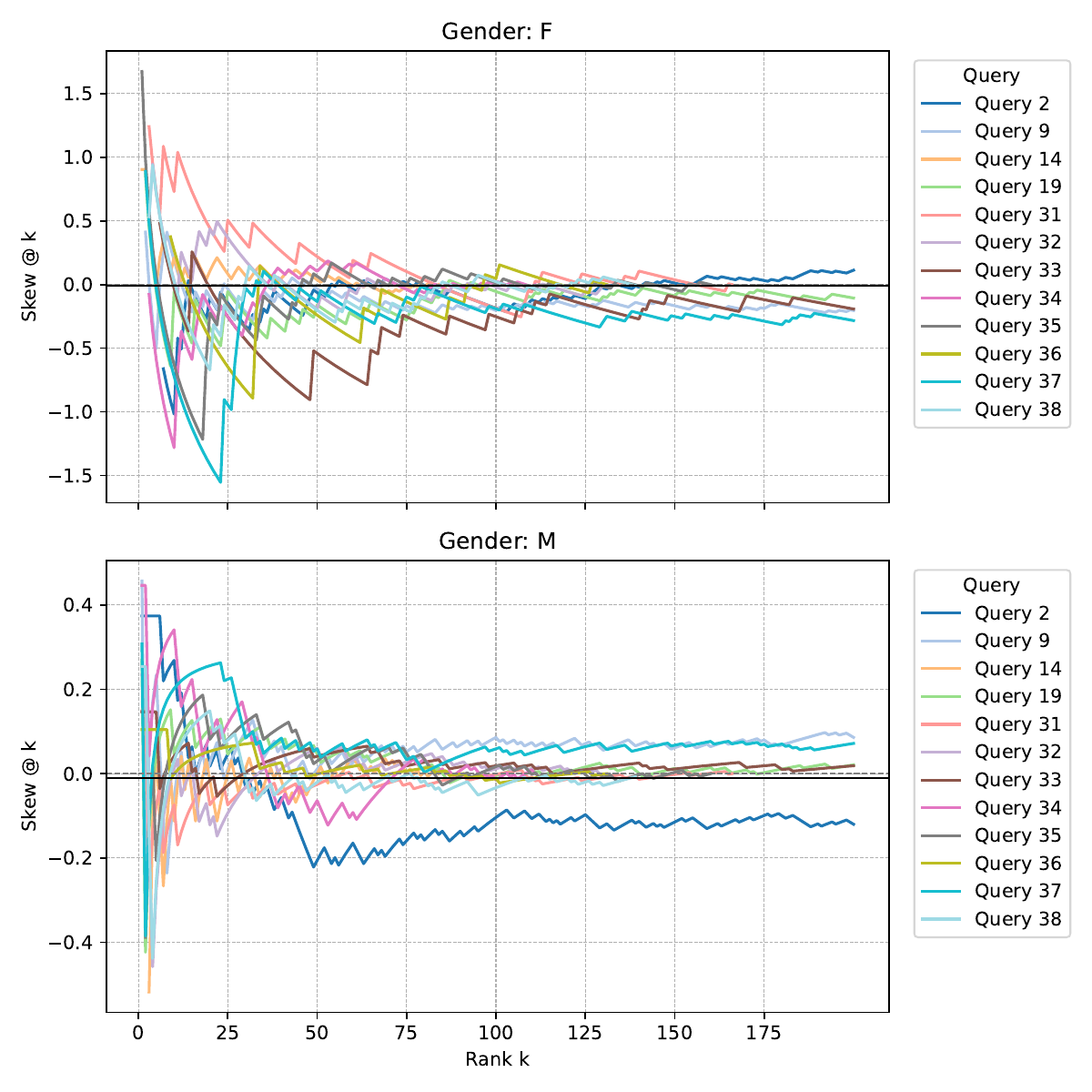}
    \caption{Skew@$k$ for each gender across 12 queries with less than 1\% missing candidates, evaluated up to rank 200. The top plot shows skew values for women; the bottom plot shows skew values for men. The black horizontal line indicates the average $\text{MinSkew}@100$ value of $-0.011$ reported in \citet{geyik2019fairness}.}
    \label{fig:skew_genders_200}
\end{figure}

The logarithmic nature of the metric makes it highly sensitive to small discrepancies in observed vs. expected proportions. While this makes it powerful for detecting even mild imbalances, it also means that inaccurate or noisy estimates of $p^*$, especially in external audits like ours, can lead to misleading skew values. Accurate estimation of the target distribution is therefore essential for the meaningful application of this metric. Therefore, for the following comparison, we focus on the subset of queries with less than 1\% missing candidates from the set where the full candidate pool was retrieved ($Q_3 \cup Q_3^\prime$). This results in 12 queries. 
Figure \ref{fig:skew_genders_200} displays the $\text{Skew}@k$ metric for both gender groups across the 12 queries, evaluated up to rank 200. The top panel shows the skew for women, while the bottom panel shows the skew for men. Note that the $y$-axis scales differ between the two plots: female skew values range from approximately $-1.5$ to $+1.5$, while male skew values are mostly bounded between $-0.4$ and $+0.4$. This indicates that deviations from expected representation are more extreme and variable for women. 

We also checked for any correlation between $\text{Skew}@k$ and group size (overall proportion) and observed no patterns (see Appendix \ref{app:results_gender}). Across queries, we observe that $\text{Skew}@k$ for women tends to be sharply negative at early ranks, indicating significant under-representation at the top of the list. For many queries, skew gradually approaches zero as $k$ increases, consistent with \textsc{DetGreedy}'s goal of aligning cumulative group representation with expected proportions. However, noticeable dips and peaks occur at rank intervals that align with page boundaries in LinkedIn Recruiter (e.g., at $k = 25, 50, 75$), which could be due to certain applied ranking-metric optimizations taking place at these cut‐offs.
 
A potential confounder in interpreting the Skew@$k$ metric at the very top of the ranking is the discrete nature of candidate placements. For small values of $k$, only a few proportions are actually attainable. For example, when $k = 3$, one can only realize 33\%, 66\%, or 100\% representation. Even in an optimally balanced ranking it might be impossible to fully eliminate skew among the early ranks of queries. 
To correct for this, we compute, at each rank cutoff $k$, the “best possible” skew that could be achieved given the group sizes and $k$, and then subtract that baseline from our observed skew. The resulting normalized deviation more accurately isolates true over- or under-representation beyond what is imposed by integral candidate placement. Our results show that even after subtracting the best attainable skew at each cutoff, the female candidates' skew at the very top ranks remains well below the reported 0.011 threshold (see Figure~\ref{fig:minskew_corrected} in Appendix~\ref{app:results_gender} for baseline‐corrected skew values). 
Furthermore, for every cutoff $k$, the female group consistently exhibits the lowest possible skew across these queries and therefore derives the MinSkew@$k$ value, a consistency that itself could indicate a potential underlying discrepancy in representation.

To assess whether the average MinSkew deviates from the reported benchmark of \(-0.011\) after accounting for both between-query variability and day-to-day noise, we fitted intercept-only linear mixed-effects models at page cutoffs \(k \in \{25,50,75,100\}\), with a random intercept for each query (see Appendix \ref{app:skew_statistical_tests} for full details). In every case through page 4 (\(k=100\)), the Wald tests reject 
\[
H_0:\;E[\text{MinSkew}@k] = -0.011
\]
with \(p<0.001\).  This confirms that the observed  MinSkew is significantly more negative than \(-0.011\) and cannot be explained by day-to-day or query-to-query noise alone.
This corroborates our findings in Section~\ref{subsec:group-deviation} which point to group disparities in the early (before 100) ranks. 
Both metrics consistently highlight that disparities are most severe in the top portion of the list and mostly in favor of the male group.


\section{Temporal Aspects of Fairness in Ranked Candidate Lists} \label{sec:temporal_analysis}
LinkedIn Recruiter operates as a two-sided marketplace, where recruiters and job seekers have distinct goals and preferences. On one side, recruiters performing repeated searches for the same role expect fresh results each day, ideally surfacing new candidates they have not yet reviewed. On the other side, candidates who are relevant to a given search would reasonably expect to appear consistently in the results across time, ensuring continued visibility.

The LTS system must balance these competing objectives.
It is essential for the ranking system to treat all demographic groups equitably in this regard. Specifically, there should be no systematic disparities in candidate retention in search results across days based on gender, race, or other protected attributes. Discrepancies in temporal exposure can lead to unequal visibility and opportunity, even if single-day rankings appear fair in isolation.

Temporal fairness, and more specifically subgroup retention across repeated queries, is an underexplored dimension in the algorithmic fairness literature \citep{patro2022fair, liu2018delayed}, and one that, despite its importance, is not explicitly mentioned in any of the LinkedIn public-facing communications.
To evaluate subgroup stability over time, we define the \emph{churn rate} for group \( g_i \) between two rankings ($r_t$) of the same query on start day \( s \) and end day \( e \), over the top \( k \) results as:

\[
\text{Churn}_{g_i}^{s \rightarrow e}(k) = 
\frac{
\left| \left\{ x \in g_i \,\middle|\, x \in \text{Top-}k(r_s) \land x \notin \text{Top-}k(r_e) \right\} \right|
}{
\left| \left\{ x \in g_i \,\middle|\, x \in \text{Top-}k(r_s) \right\} \right|
}
\]

This metric captures the proportion of candidates in group \( g_i \) who appeared in the top \( k \) results at time \( s \) but were no longer present at time \( e \). While the overall level of churn, whether high or low, may reflect deliberate design choices (e.g., promoting freshness vs.~stability), it is important for the churn rate to be approximately consistent across demographic groups. This is due to the fact that large disparities in churn can lead to unequal exposure over time, which in turn can create disparate candidate outcomes.

\begin{figure}[h]
    \centering
    \includegraphics[scale=0.35]{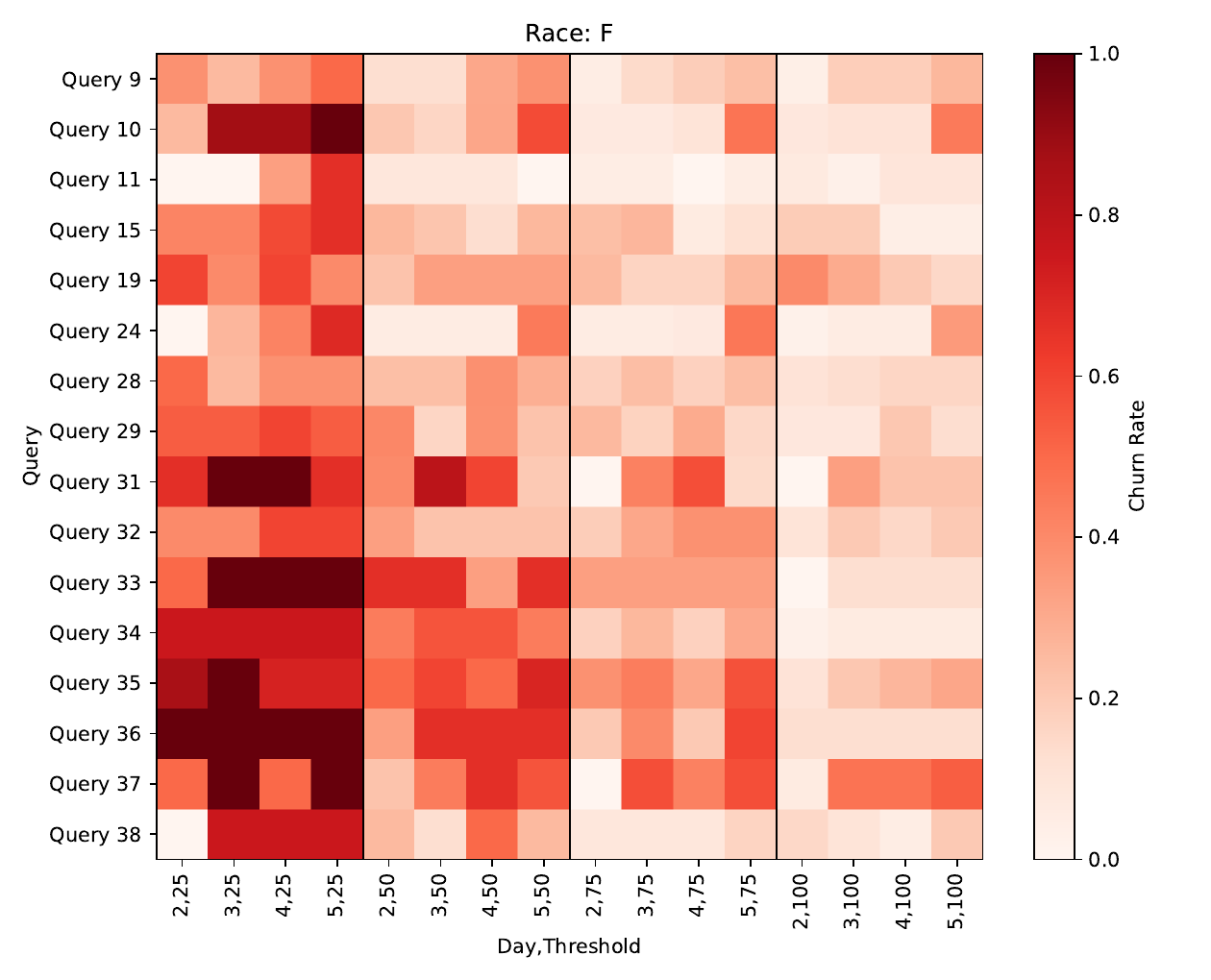}
    \includegraphics[scale=0.35]{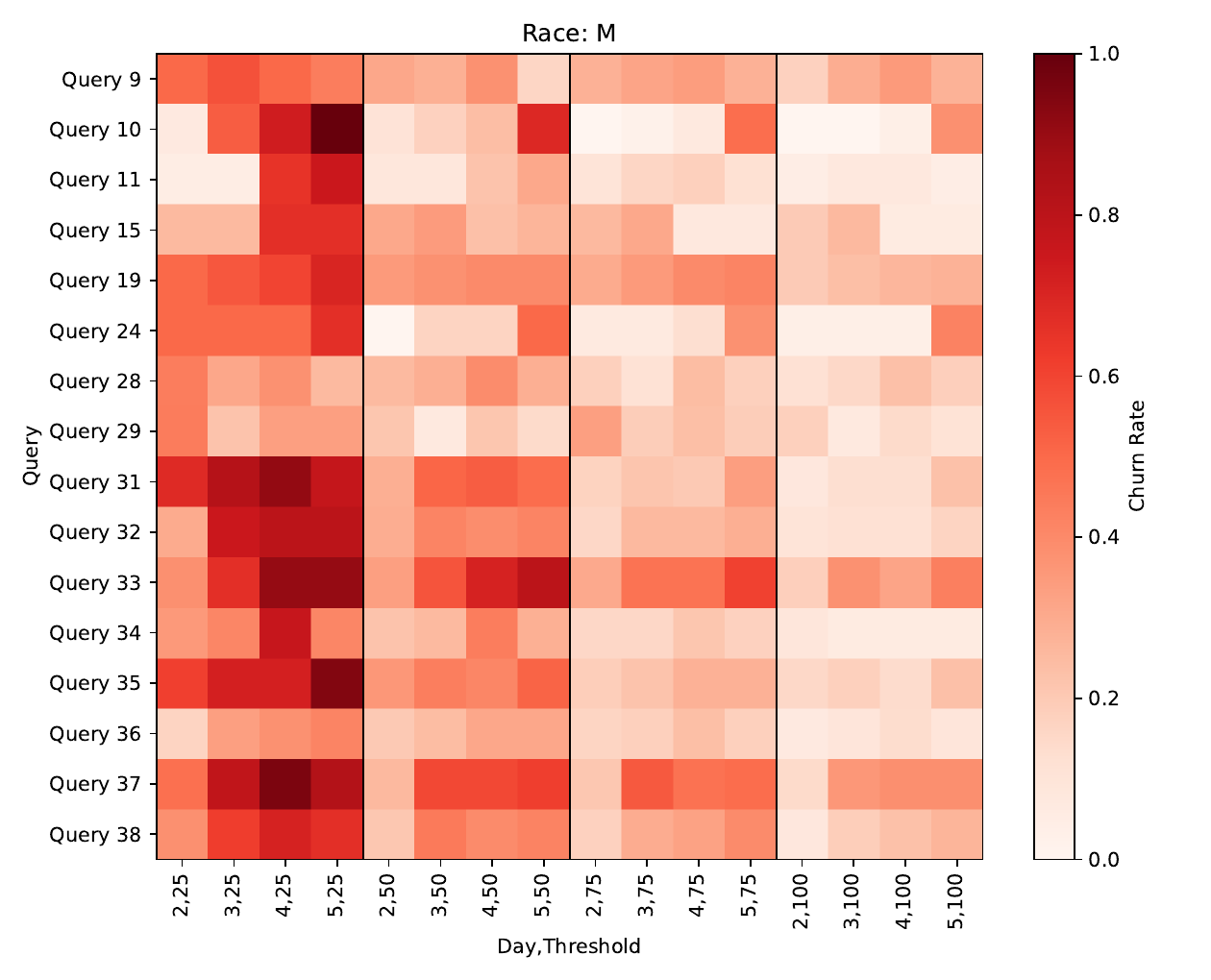}
    \caption{(\textbf{Left}): Heat map of day‐to‐day churn rates \(\mathrm{Churn}_{g_i}^{1\rightarrow j}(k)\) for the female candidates across position‐specific queries (with less than 15\% missing candidates) over five consecutive days, evaluated at top‐\(k\) cutoffs \(k\in\{25,50,\dots,200\}\). Darker shades indicate higher proportions of candidates dropping out of the top \(k\).
    (\textbf{Right}): Identical plot for male candidates.}
    \label{fig:churn_rate_specific_F}
    \vspace{-10pt}
\end{figure}

The left and right panels of Figure~\ref{fig:churn_rate_specific_F}, for females and males respectively, show $\text{Churn}_{g_i}^{1 \rightarrow j}(k)$ for $k \in \{25, 50, \ldots, 200\}$ and $j \in \{2,3,4,5\}$ in the set $Q_3 \cup Q'_3$ of position specific queries  with less than 15\% missing candidates and 5 consecutive days of data.
Higher churn rate, indicating a greater proportion of individuals dropping out of the top-$k$ ranking, is represented using a darker shade.
Recall, though, that our auditing pipeline does not allow us to determine whether departures are due to hires or algorithmic reshuffling.

Overall, we observe that churn rates are highest at the very top of the rankings and then steadily decline as $k$ increases, reflecting the competition for top slots and a relative stabilization deeper in the list. This decrease with larger cut-offs holds for both genders.

Notably, at $k=25$ and $k=50$ women churn approximately 0.07 units more than men on average across all days, indicating a less stable presence in the top-$k$ pools. 
Furthermore, male drop-outs follow a more predictable pattern; turnover from day 1$\rightarrow$2 is smaller than from day 1$\rightarrow$3, 4, or 5, whereas women's exits appear more erratic, suggesting additional volatility in their rankings. 
We examined whether churn rates correlate with each group’s overall representation to explain the observed discrepancies and found no correlation between group proportions and churn rates.
We further conducted a Wald test via mixed‐effects models to test for significant group differences and observed statistically significant differences in churn at $k=25$ and $k=50$ between the two groups.

We also provide a parallel churn‐rate analysis, both for the position‐specific queries across racial groups and for our set of general queries.
Precise results and additional methodological details appear in Appendix~\ref{app:results_temporal}.

\section{Limitations} \label{app:limitations}
Our study has several important limitations that should be considered when interpreting its results. First, the query distribution we utilize may not precisely reflect the true distribution of queries issued by LinkedIn recruiters in practice. Although any deterministic post-processing or ranking algorithm is likely to apply uniformly to all non-sensitive queries, our reliance on static query snapshots does not fully capture the interactive and iterative nature of recruiter workflows, where recruiters often refine their searches after initial results and interactions with candidates. Despite this, analyzing static queries remains valuable for understanding the fundamental behavior of LinkedIn's ranking algorithms.

Moreover, our methodology faces constraints related to data collection and candidate inference. Our demographic inference, particularly racial inference, introduces substantial noise that limits the reliability of fine-grained metrics such as MinSkew. Furthermore, practical challenges prevented us from investigating certain ranking behaviors through controlled experimentation with synthetic candidate profiles: attempts to create these profiles were hindered by LinkedIn's automated spam detection and identity verification processes, raising feasibility and ethical concerns. Finally, our dataset represents only a small sample of the extensive LinkedIn candidate network, and thus our observations may not generalize to the platform as a whole or over longer periods of observation.

These suggest the following concrete capabilities that LinkedIn should give to independent auditors in the public interest.
\begin{itemize}
    \item Provide API access to LinkedIn Talent Search under a fair use model that does not depend on personal account privileges or restrictive subscription tiers.
    \item Increase transparency around deployed ranking and post-processing methods and interventions through technical documentation or model cards. This would offer critical context for interpreting observed patterns, such as the unexplained behavior at page cut-offs in our study, and improve the interpretability of audit results.
    \item Enable privacy-preserving access to demographic labels or release more granular aggregated demographic data, such as group distributions at the state or metropolitan level, to improve baseline comparisons in fairness assessments.
    \item Enable researchers to create sandboxed synthetic accounts that do not appear in real recruiter searches but can be retrieved using a designated key or testing mode. These accounts would allow researchers to conduct more controlled experiments to analyze the behavior of the ranking algorithm under different interventions.
\end{itemize}
Enabling these capabilities would significantly improve the feasibility, rigor, and reproducibility of third-party audits, while upholding platform accountability and fairness.

\section{Conclusion}
We performed an independent external audit of LinkedIn Recruiter's ranking algorithms, assessing potential disparities in candidate representation across gender and racial groups. Our analysis indicates that LinkedIn likely employs some form of demographic-aware post-processing, evident from diminishing disparities deeper in the rankings. However, this mechanism appears less effective at the highest ranks, where representation remains notably imbalanced, highlighting a critical area for improvement. We further found evidence of temporal instability in candidate rankings, with churn rates differing across groups. Beyond these specific observations, our work provides methodological insights that contribute to the broader understanding of how independent algorithmic audits can be systematically carried out on platforms with restricted access. We discuss the limitations of our audit in Section~\ref{app:limitations}; importantly, the limitations can inform transparency desiderata to facilitate meaningful auditing.

\subsection*{Acknowledgments}
Our study received IRB exemption UP-24-01124 at the University of Southern California.
This work was supported by multiple National Science Foundation grants: CNS-$1956435$, CNS-$2344925$, and NSF CAREER Award CCF-$2239265$.
V. Sharan was also supported by an Okawa Foundation Award, and A. Korolova by the
Alfred P. Sloan Research Fellowship.

We thank the anonymous AAAI reviewers for helpful comments and feedback used to improve the work.

\bibliography{aaai2026}
\bibliographystyle{abbrvnat}

\newpage
\appendix

\clearpage

\section{Full Query Information}
\subsection{General Queries} \label{app:general_queries}
Table \ref{tab:query_info_geneeral_queries} shows a list of queries scraped up to rank 200 and the number of missing candidates for each of these general queries on the first day.
\begin{table}[h!]
\small
    \centering
    \begin{tabular}{c|c}
    \toprule
         Query& Missing Number  \\
         \midrule
        Chiropractor&41\\
        Civil engineer&4\\
        Operations research analyst&3\\
        Pharmacist&39\\
        Public relations specialist&5\\
        Software developer&0\\
        Special education teacher&44\\
        Tutor&18\\
        Aerospace engineer&27\\
        Carpenter&125\\
        Chemical engineer&24\\
        Computer network architect&6\\
        Graphic designer&10\\
        Industrial engineer&6\\
        Interior designer&22\\
        translator&51\\
        Computer hardware engineer&14\\
        Editor&7\\
        Financial analyst&6\\
        Financial manager&4\\
        Marketing manager&1\\
        Materials engineer&6\\
        Paralegal&4\\
        Retail salesperson&93\\
        Chief executive&4\\
        Compliance officer&4\\
        Construction manager&16\\
        Cost estimator&18\\
        Fundraiser&4\\
        Human resources manager&2\\
        Insurance underwriter&22\\
        Project management specialist&4\\
        Architect&10\\
        Computer programmer&0\\
        Database administrator&44\\
        Electrical engineer&23\\
        Information security analyst&18\\
        Mathematical scientist&32\\
        Tax preparer&42\\
        Web developer&10\\
        \bottomrule
    \end{tabular}
    \caption{General queries with results scraped up to rank 200 (first eight pages) and the number of missing candidates per query. See Section \ref{sec:data_retrieval} for details.}
    \label{tab:query_info_geneeral_queries}
\end{table}
\clearpage

\subsection{Position Specific Queries} 
\label{app:position_specific_queries}
\begin{description}[
    style=nextline,
    leftmargin=*,
    labelwidth=3em,
]
\tiny
    \item[Query 1]
    \texttt{``Process Engineer'' AND (``renewable fuels'' OR biofuels OR biodiesel OR ``green hydrogen'') AND (``process design'' OR ``process optimization'' OR ``scale-up'') AND (``Aspen Plus'' OR HYSYS OR ChemCAD) AND sustainability AND (senior OR ``mid-senior'')}
  \item[Query 2]
    \texttt{``Research Scientist'' AND AI AND PhD AND (``scientific publications'' OR ``conference submissions'') AND (``Python'' OR ``JavaScript'' OR R OR Java OR ``C++'')}
  \item[Query 3]
    \texttt{(``5th grade teacher'' OR ``elementary teacher'' OR ``upper elementary educator'') AND (``culturally and linguistically diverse'' OR ``CLD endorsement'' OR ``ELL support'')}
  \item[Query 4]
    \texttt{(``AML compliance analyst'' OR ``anti-money laundering analyst'' OR ``transaction monitoring analyst'') AND (``sanctions screening'' OR ``suspicious activity reports'' OR ``SARs'') AND (``FinCEN 314(a)'' OR ``regulatory compliance'') AND (``financial services'' OR ``banking industry'') AND (``AML compliance'' AND ``transaction monitoring'') AND (ACAMS OR ``certified anti-money laundering specialist'')}
  \item[Query 5]
    \texttt{(``Actuation Bench Technician'' OR ``Actuation Technician'' OR ``Bench Technician'') AND (mechanical OR hydraulic OR pneumatic OR electrical) AND (``diagnostic tools'' OR ``testing equipment'')}
  \item[Query 6]
    \texttt{(``Eligibility Administrator'' OR ``Eligibility Specialist'') AND (``health insurance'' OR ``medical insurance'' OR ``group plans'') AND (``eligibility processing'' OR enrollment OR ``data entry'')}
  \item[Query 7]
    \texttt{(``Graphic Designer'' OR ``Visual Designer'') AND (``Adobe Photoshop'' OR Illustrator OR InDesign) AND (``print-ready'' OR packaging OR ``marketing materials'') AND (Nintendo OR ``interactive entertainment'')}
  \item[Query 8]
    \texttt{(``Talent Coordinator'' OR ``Interview Scheduler'') AND (``candidate experience'' OR ``talent acquisition'') AND (``calendar management'' OR ``interview logistics'') AND (``ATS'' OR ``applicant tracking'')}
  \item[Query 9]
    \texttt{(``advanced analytics senior associate'' OR ``senior data analyst'' OR ``analytics consultant'') AND (``SQL'' OR ``Python'' OR ``applied statistics'') AND (``A/B testing'' OR ``CRM data'' OR ``customer operations analytics'') AND (``business intelligence'' AND ``big data analytics'') AND (``LinkedIn'' OR ``advertising data'' OR ``customer support analytics'')}
  \item[Query 10]
    \texttt{(``agile scrum master'' OR ``scrum lead'' OR ``scrum master'' OR ``agile program lead'') AND (``product owner experience'' OR ``de facto product owner'') AND (``senior'' OR ``mid-level'')}
  \item[Query 11]
    \texttt{(``amazon business analyst'' OR ``ecommerce analyst'' OR ``CPG analyst'' OR ``digital commerce analyst'') AND (``ecommerce metrics'' OR ``Amazon Seller Central'' OR ``performance reporting'' OR ``DTC analytics'')}
  \item[Query 12]
    \texttt{(``architecture designer'' OR ``architectural designer'' OR ``architectural associate'') AND (``building design'' AND ``schematic design'' OR ``construction documents'') AND (``Revit'' AND ``Adobe Creative Suite'') AND (``building codes'' OR ``ADA compliance'' OR ``zoning regulations'') AND (``NCARB'' OR ``AXP'' OR ``architectural registration'')}
  \item[Query 13]
    \texttt{(``assistant fashion editor'' OR ``fashion assistant'' OR ``fashion coordinator'') AND (``fashion market experience'' OR ``sample management'' OR ``fashion shoots logistics'')}
  \item[Query 14]
    \texttt{(``business development manager'') AND (``A/E industry'' OR ``architecture engineering'' OR ``industrial projects'') AND (``industrial manufacturing'' OR ``design-build'' OR ``engineering firm'') AND (``client relationships'' OR ``market strategy'' OR ``capture planning'') AND (``CRM'' OR ``Vantagepoint'')}
  \item[Query 15]
    \texttt{(``clinical scientist'' OR ``medical monitor'' OR ``clinical research scientist'') AND (``medical monitoring'' OR ``safety review'' OR ``clinical data review'' OR ``protocol adherence'') AND (GCP OR ICH OR ``clinical trials'' OR IND OR ``regulatory submissions'') AND (PhD OR PharmD OR ``Nurse Practitioner'' OR NP OR PA) AND (``neuromuscular'' OR ``rare disease'' OR ``CNS'')}
  \item[Query 16]
    \texttt{(``creative director'' OR ``senior art director'' OR ``head of creative'') AND (``packaging design'' OR ``retail branding'' OR ``store experience'') AND (``CPG'' OR ``beauty'' OR ``wellness'' OR ``lifestyle'' OR ``fashion'') AND (``Adobe Creative Suite'' OR Photoshop OR Illustrator OR InDesign) AND (``team leadership'' OR ``content production'' OR ``store design'') NOT (freelance OR intern OR assistant OR ``entry level'') AND (``creative strategy'' OR ``brand refresh'' OR ``content production'')}
  \item[Query 17]
    \texttt{(``criminal investigator'' OR ``protective service agent'' OR ``executive protection officer'') AND (``GS-1811'' OR ``federal law enforcement'' OR ``protective services training'' OR ``FLETC'') AND (``executive protection'' OR ``protective detail'' OR ``threat assessment'' OR ``liaison with law enforcement'') AND (``Top Secret clearance'' OR ``firearms qualification'' OR ``site evaluation'' OR ``travel protection'')}
  \item[Query 18]
    \texttt{(``cybersecurity analyst'') AND (``vulnerability management'' OR ``SPLUNK'') AND (``Microsoft SCCM'' OR ``PowerShell'') AND (``NIST 800-53'' OR ``CIS Controls'')}
  \item[Query 19]
    \texttt{(``data products analyst'' OR ``analytics engineer'') AND (``SQL'' OR ``Spark'') AND (``ads data'' OR ``business intelligence products'' OR ``reporting systems'') AND (senior OR ``mid-level'')}
  \item[Query 20]
    \texttt{(``director of hardware'' OR ``field service director'' OR ``hardware operations manager'') AND (``field service operations'' OR ``installation planning'' OR ``hardware deployment'')}
  \item[Query 21]
    \texttt{(``director of operations'' OR ``internal operations director'' OR ``business operations leader'' OR ``operations executive'') AND (``utilities'' OR ``infrastructure'' OR ``energy'' OR ``real estate technology'' OR ``B2B services'') AND (``billing operations'' OR ``recurring revenue'' OR ``submetering'' OR ``SaaS operations'')}
  \item[Query 22]
    \texttt{(``laboratory technician'' OR ``lab tech'') AND (``quality control'' OR ``product inspection'' OR ``standard operating procedures'') AND (``chemical manufacturing'' OR ``industrial laboratory'' OR ``rotating shift'') AND (``hazardous waste handling'' OR ``safety protocols'')}
  \item[Query 23]
    \texttt{(``litigation paralegal'' OR ``defense paralegal'') AND (``personal injury'' OR ``medical litigation'') AND (``drafting pleadings'' OR ``discovery responses'') AND (``PACER'' OR ``NYSCEF'')}
  \item[Query 24]
    \texttt{(``marketing director'' OR ``head of marketing'' OR ``brand director'') AND (``restaurant industry'' OR ``foodservice marketing'' OR ``hospitality'') AND (``social media strategy'' OR ``loyalty programs'') AND (``campaign management'' OR ``brand strategy'') AND (``team leadership'' OR ``budget ownership'') AND (senior OR ``mid-level'')}
  \item[Query 25]
    \texttt{(``medical review analyst'' OR ``clinical review nurse'' OR ``utilization review nurse'') AND (``medical necessity'' OR ``benefit eligibility'' OR ``claims review'' OR ``reimbursement analysis'') AND (``claims determination'' OR ``clinical review'' OR ``coding review'')}
  \item[Query 26]
    \texttt{(``news writer'' OR ``broadcast writer'' OR ``script writer'' OR ``news copywriter'') AND (``international news'' OR ``foreign affairs'') NOT (intern OR internship OR student OR freelance)}
  \item[Query 27]
    \texttt{(``roadway design'' OR ``site grading'' OR ``drainage'' OR ``municipal engineering'') AND (AutoCAD OR ``civil 3d'') AND (``Bachelor's in Civil Engineering'' OR ``BSCE'')}
  \item[Query 28]
    \texttt{(``strategy manager'' OR ``business strategy manager'' OR ``corporate strategy associate'') AND (``retail strategy'' OR ``consumer goods strategy'' OR ``apparel industry strategy'') AND (``market expansion'' OR ``customer acquisition'' OR ``business development'' OR ``competitive intelligence'')}
  \item[Query 29]
    \texttt{(``talent development specialist'' OR ``learning and development specialist'' OR ``organizational development specialist'') AND (``learning program design'' OR ``leadership development'' OR ``employee development initiatives'') AND (``LMS management'' OR ``learning management systems'' OR ``training delivery'') AND (``training facilitation'' OR ``enterprise learning solutions'' OR ``continuous learning programs'')}
  \item[Query 30]
    \texttt{(title:``Architectural Designer'' OR title:``Project Designer'') AND (Revit AND AutoCAD) AND (``Bachelor of Architecture'' OR ``Master of Architecture'') AND (``3+ years'' OR ``3 years'' OR experience)}
    
  \item[Query 31]
    \texttt{(``software engineer'' OR ``cloud software engineer'') AND (``distributed systems'' OR ``large-scale system design'') AND (``Python'' AND ``Java'' AND ``C++'') AND (``Google Cloud'' AND ``cloud computing'') AND (``master's degree'' OR ``PhD'') AND (``data structures'' AND ``algorithms'')}
  \item[Query 32]
    \texttt{(``interaction designer'' OR ``senior UX designer'' OR ``senior product designer'') AND (``mobile design'' OR ``desktop design'' OR ``cross-platform'') AND (``consumer app design'' OR ``mobile product design'')}
  \item[Query 33]
    \texttt{(``research scientist'' OR ``AI research scientist'') AND (``generative AI agents'' OR ``LLM-based agents'' OR ``reinforcement learning agents'')}
  \item[Query 34]
    \texttt{(``distributed systems engineer'' OR ``senior backend engineer'' OR ``platform engineer'') AND ((``Java'' OR ``C\#'') AND ``object-oriented programming'') AND (``Kafka'' OR ``Cassandra'' OR ``Elasticsearch'' OR ``AWS'') AND (``gRPC'' OR ``GraphQL'') AND (``high availability'' OR ``fault tolerance'')}
  \item[Query 35]
    \texttt{(``frontend engineering manager'' OR ``UI engineering manager'' OR ``web engineering lead'') AND (``accessibility'' OR ``design systems'' OR ``developer experience'')}
  \item[Query 36]
    \texttt{(``senior staff software engineer'' OR ``principal software engineer'' OR ``staff software engineer'') AND (``payment systems'' OR ``financial platforms'' OR ``payment workflows'') AND (``software architecture'' AND ``system design'' OR ``project leadership'')}
  \item[Query 37]
    \texttt{(``software engineer'' OR ``machine learning engineer'') AND (``recommendation systems'') AND (``model deployment'' OR ``ML infrastructure'' AND ``optimization'') AND (``data structures'' AND ``algorithms'') AND (``Python'' AND ``C++'' AND ``Java'')}
  \item[Query 38]
    \texttt{(``full stack engineer'' OR ``senior software engineer'') AND (``Java'' OR ``Swift'' OR ``Objective-C'' OR ``JavaScript'') AND (``mobile development'' OR (``iOS'' AND ``Android'')) AND (``APIs'' OR ``SDKs'' OR ``ad platform'' OR ``interactive ads'') AND (``Google Cloud Platform'' OR ``Kubernetes'' OR ``BigQuery'') AND (``SSP'' OR ``DSP'' OR ``VAST'' OR ``Ad Serving'')}
\end{description}

Table \ref{tab:query_info_specific_queries} summarizes, for each position-specific query, the number of missing candidates and the total size of the returned pool. $Q_3$ includes queries 1--30, and queries 31--38 comprise $Q'_3$. 
\begin{table}[ht!]
\centering
\begin{tabular}{lrr}
\toprule
Query & Maximum Rank & Missing Numbers\\
\midrule
Query 1  &  931 &  16\\
Query 2  &  571 &   6\\
Query 3  &  211 &  88\\
Query 4  &  320 &  11\\
Query 5  &  152 &  73\\
Query 6  &  993 & 541\\
Query 7  & 1000 & 100\\
Query 8  &  205 &   5\\
Query 9  &  448 &   1\\
Query 10 &  832 &  24\\
Query 11 &  591 &  41\\
Query 12 &  271 &  19\\
Query 13 &  431 &  51\\
Query 14 &   84 &   0\\
Query 15 &  218 &   3\\
Query 16 & 1000 &  34\\
Query 17 &   92 &  17\\
Query 18 &  946 &  67\\
Query 19 &  603 &   2\\
Query 20 &   72 &  10\\
Query 21 &  558 &  12\\
Query 22 &  761 & 240\\
Query 23 &  818 & 223\\
Query 24 &  731 &   9\\
Query 25 &  781 & 428\\
Query 26 &  624 & 101\\
Query 27 &  750 & 107\\
Query 28 &  810 &   9\\
Query 29 &  462 &  24\\
Query 30 &  520 &  50\\
Query 31 &  166 &   0\\
Query 32 &  130 &   1\\
Query 33 &  666 &   7\\
Query 34 &  111 &   0\\
Query 35 &  160 &   0\\
Query 36 &  131 &   0\\
Query 37 &  697 &   2\\
Query 38 &  139 &   1\\
\bottomrule
\end{tabular}
\caption{Missing candidates and maximum rank (number of available candidates in total for that query) for position-specific queries.}
\label{tab:query_info_specific_queries}
\end{table}

\section{Additional Results and Analyses}
\label{app:results}
In this section, we extend the results from Sections~\ref{sec:exposure_disparity_analysis} and \ref{sec:temporal_analysis} by applying the same analyses to our set of general queries ($Q_2$) and to racial groups. 

\subsection{Auditing for Gender-Based Post-Processing} \label{app:results_gender}

In this section, we aim to calculate the deviation from mean group proportions for our set of general queries ($Q_2$). We pursue an alternative approach to estimate $p_i^*$ in the absence of the entire candidate pool, compare this approach to using observed proportions, and discuss limitations of both methods.
One difficulty with general queries is that we cannot reliably estimate $p_i^*$ from the data alone. To address this, we use group proportions reported in the U.S.~Bureau of Labor Statistics' (BLS) 2024 data corresponding to each query's occupational category.

\begin{figure}[h!]
    \centering
    \includegraphics[width=1\linewidth]{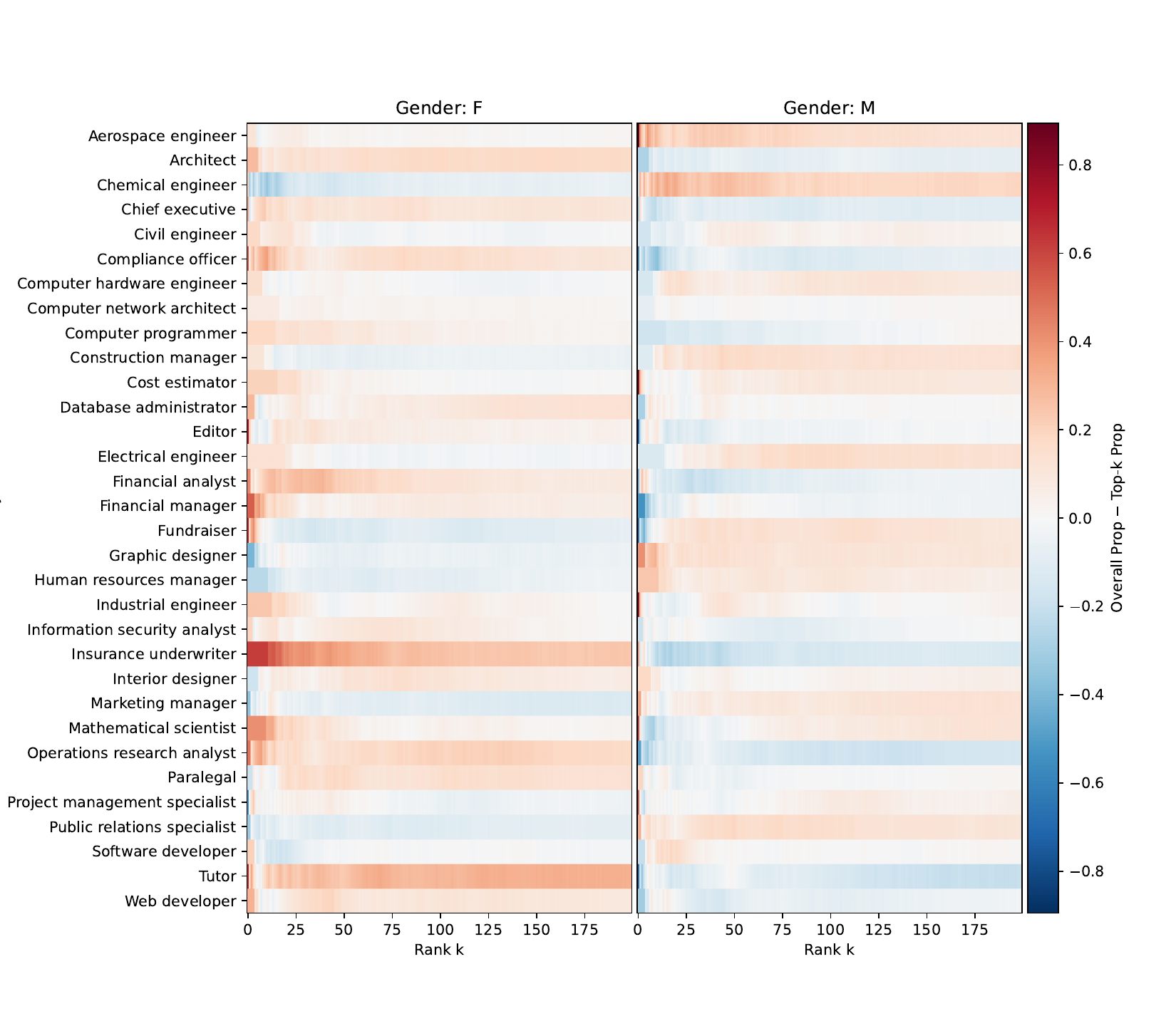}
    \caption{Deviation between the overall BLS-based gender proportion and the top-$k$ proportion at each rank position $k$, across all queries. The left panel shows deviations for female candidates, and the right panel shows deviations for male candidates. Positive values (red) indicate under-representation in the top ranks relative to the overall group proportion for that query, while negative values (blue) indicate over-representation. Each row corresponds to a different query, labeled by its occupational title.}
    \label{fig:gender_deviation_BLS}
\end{figure}

\begin{figure}[h!]
    \centering
    \includegraphics[width=1\linewidth]{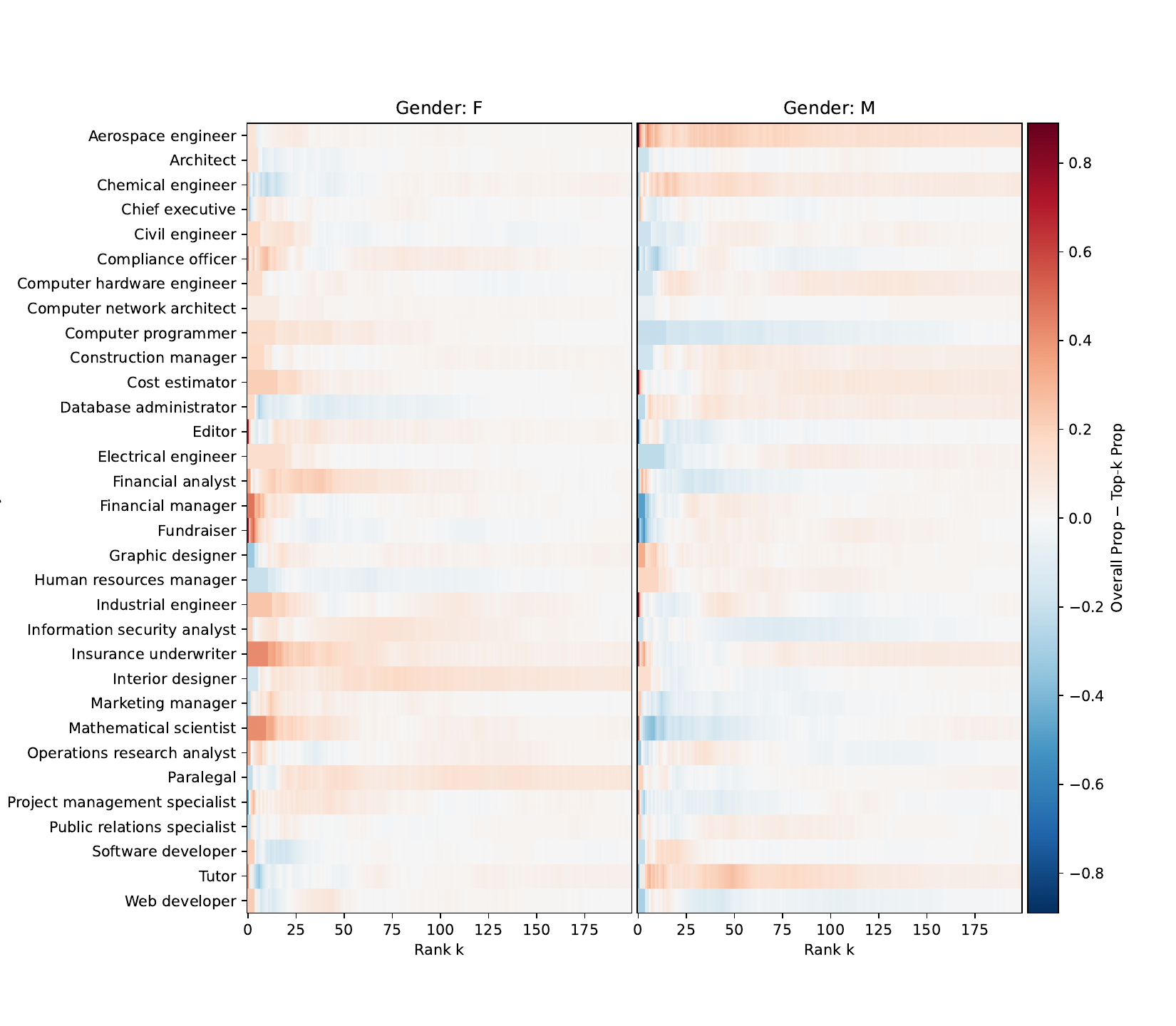}
    \caption{Deviation between the overall gender proportion in the top 200 candidates and the top-$k$ proportion at each rank position $k$, across all queries. The left panel shows deviations for female candidates, and the right panel shows deviations for male candidates. Positive values (red) indicate under-representation in the top ranks relative to the overall group proportion for that query, while negative values (blue) indicate over-representation.}
    \label{fig:gender_deviation_actual_value}
\end{figure}

Figure \ref{fig:gender_deviation_BLS} visualizes the deviation $p^*_i - \dfrac{n_{i,k}}{k}$, where $n_{i,k}$ denotes the cumulative count of candidates from group $i$ up to rank $k$, and $p^*_i$ is based on BLS data. For comparison, Figure \ref{fig:gender_deviation_actual_value} presents the same analysis using the actual group proportions among the top 200 retrieved candidates as the value of $p^*_i$. For this analysis, we excluded any queries in which more than 15\% of the top 200 results consisted of LinkedIn members without observable profile data. 

We observe persistent under-representation of women in the early ranks for many technical roles, including software developer, civil engineer, and construction manager. In contrast, male candidates are underrepresented in roles such as human resources manager and interior designer, especially in early ranks. While these deviations tend to narrow as $k$ increases, representation in the top ranks remains a critical concern, as these early positions receive the majority of recruiter attention and drive downstream outcomes.

When we use BLS proportions as our baseline, the resulting deviations are markedly larger than those computed from the observed top-200 candidate proportions. Each baseline has its own limitations. BLS figures reflect the entire U.S. workforce rather than only that of the NYC metropolitan area of our queries.\footnote{We were not able to find fine-grained, region-specific occupational demographics.} 
As a result, national proportions may not fully reflect the candidate pool composition in our setting. On the other hand, the top-200 proportions are not computed over the entire candidate pool. Therefore, we refrain from making strong claims about absolute bias magnitudes. Nonetheless, it is reassuring that both approaches yield parallel patterns, suggesting that our core findings are robust to the choice of $p_i^*$ estimation.

\clearpage
\subsubsection{Skew@$k$}

\begin{figure}[h!]
    \centering
    \includegraphics[width=0.49\linewidth]{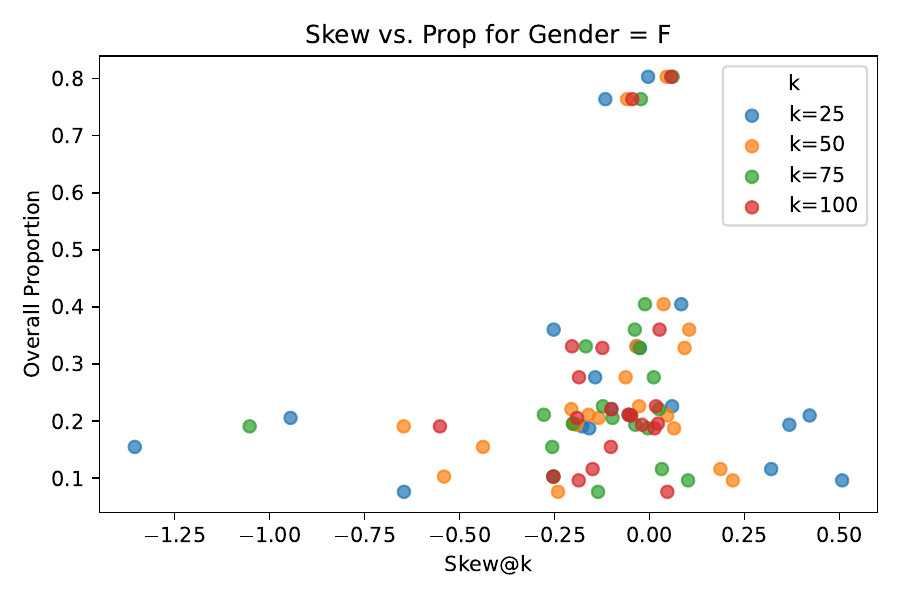}
    \includegraphics[width=0.49\linewidth]{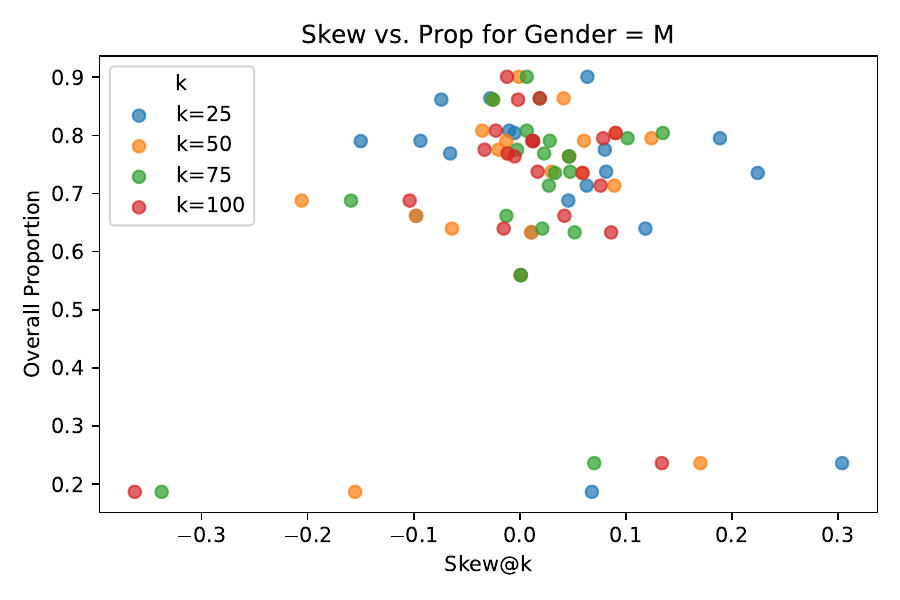}
    \caption{(\textbf{Left}): Scatter of skew@$k$ versus overall female proportion for $k=25,50,75,100$. Each color denotes a different cutoff. (\textbf{Right}): Same plot for male proportion.}
    \label{fig:skew_prop_F}
\end{figure}

\begin{figure}[h]
    \centering
    \includegraphics[scale=0.5]{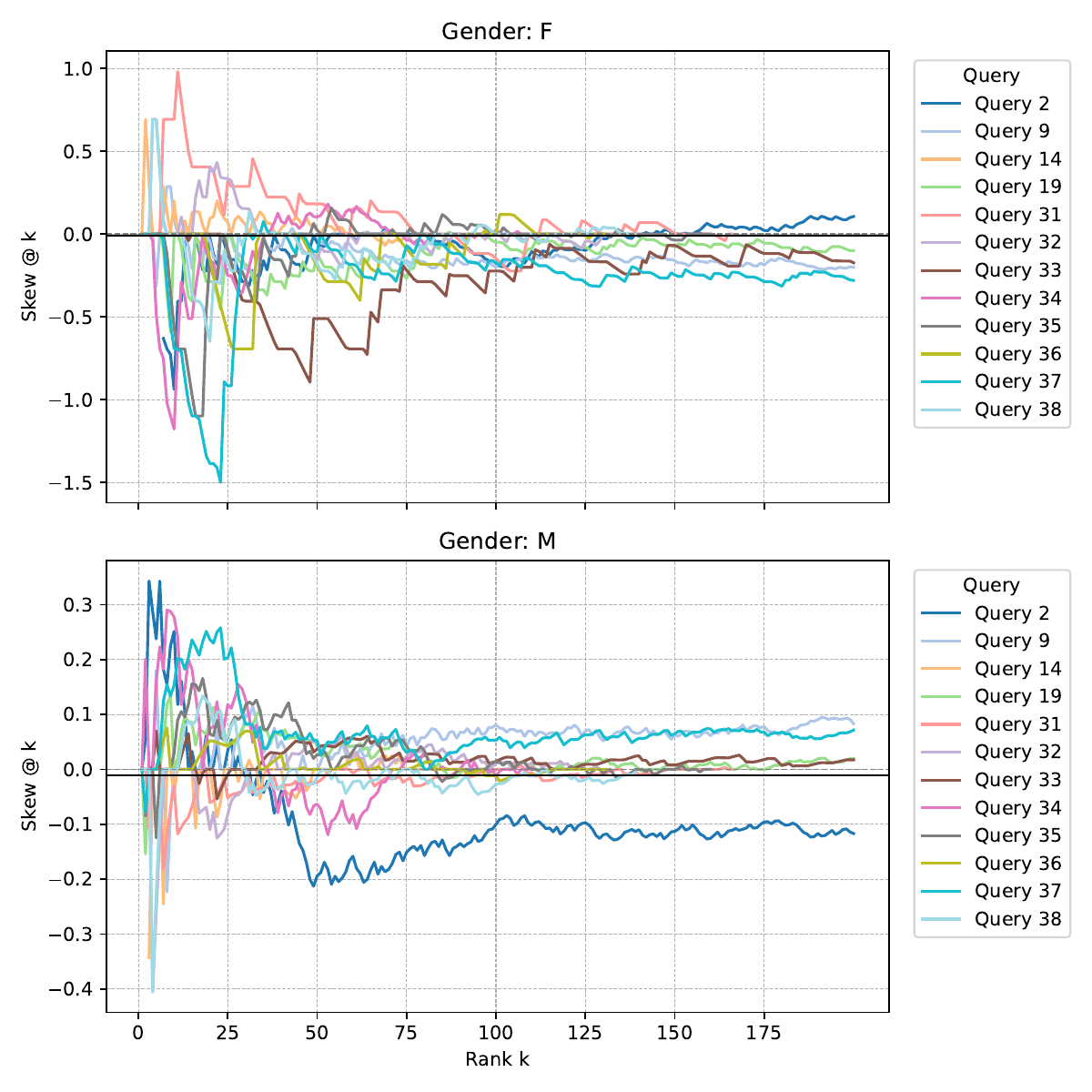}
    \caption{Integrality-corrected skew for each gender across 12 queries (with less than 1\% missing candidates), evaluated up to rank 200.}
    \label{fig:minskew_corrected}
\end{figure}

In Section \ref{sec:result_skew}, we reviewed the definition of MinSkew@$k$ and reported on its values for queries in $Q_3$ and $Q'_3$. 
In this section, we begin by examining whether group size might be driving observed disparities; we find no meaningful correlation between MinSkew@$k$ and overall group proportions (Figure~\ref{fig:skew_prop_F}). To account for the discrete nature of candidate rankings, we then normalize the skew values by subtracting the best attainable skew for each group and observe that the overall patterns remain unchanged (Figure \ref{fig:minskew_corrected}).

In order to verify that the magnitude of skew@k is not simply driven by how large or small a group is within each query, we created scatter plots of skew@$k$ against the group’s overall proportion (group size) for four cutoffs $k=25,50,75,100$ (Figure~\ref{fig:skew_prop_F}). Each point represents one query observation, colored by the value of $k$. We do not observe any particular correlation between the two variables in any of our gender groups.

As noted in Section \ref{sec:result_skew}, the discrete nature of candidate placements can bias interpretation of the Skew@\(k\) metric. To correct for this, we compute, for each cutoff \(k\) and group \(g_i\), the unavoidable skew from integrality. Let
\[
m_i^{-}(k) = \bigl\lfloor k\,p_i^* \bigr\rfloor /k,
\qquad
m_i^{+}(k) = \bigl\lceil k\,p_i^* \bigr\rceil /k.
\]
where \(p_i^*\) is the group’s overall proportion. The unavoidable skew is then
\[
S_{g_i}^{\mathrm{int}}(k)
= \min\ \left(
  \left| \log\!\frac{m_i^{-}(k)}{p_i^*} \right|,\;
\left| \log\!\frac{m_i^{+}(k)}{p_i^*} \right|
\right).
\]

Then, for the observed skew
\[
    S_{g_i}^{\mathrm{obs}}(k) = \log \left( \frac{p^{\tau_r}_{k,{g_i}}}{p_i^*} \right),
\]
 we define a signed, integrality-corrected skew as
 \[
\widetilde{S}_{g_i}(k) = \operatorname{sign} \left( S^{\text{obs}}_{g_i}(k) \right) \cdot \left( \left| S^{\text{obs}}_{g_i}(k) \right| - S^{\text{int}}_{g_i}(k)\right).
\]
The observed skew $S^{\text{obs}}_{g_i}(k)$ reflects how far the group’s representation deviates from perfect proportionality at rank cutoff $k$, measured in log-space. However, due to integrality constraints, since only whole candidates can be selected, not all proportions are achievable. 
We define $S^{\text{int}}_{g_i}(k)$ as the minimum skew that could be attained under these constraints, even in an ideally fair ranking.

To isolate the unfairness introduced by the actual ranking (beyond what is structurally unavoidable), we subtract this baseline.
Specifically, we compute the corrected skew $\widetilde{S}_{g_i}(k)$ by removing the integrality-induced minimum from the observed skew's magnitude, while preserving its sign. Figure~\ref{fig:minskew_corrected} shows the corrected skew using the above method. The patterns for each query match those of Figure~\ref{fig:skew_genders_200}, demonstrating that even at top ranks, better (i.e., higher) skew can be achieved.

\subsubsection{Statistical Analysis}\label{app:skew_statistical_tests}
To assess whether or not at page cut-offs, the average observed MinSkew value is different from the reported average value by LinkedIn, we calculate MinSkew@$k$ for $k \in \{25, 50, 75, 100\}$ for all queries $q$ with less than 15\% missing candidates and more than 100 total candidates over all available days $d$. With 328 MinSkew observations across 72 queries (ranging from 1 day per query to 9 days per query), we fit an intercept-only linear mixed-effects model:
\[
  \text{MinSkew@}k_{q,d} = \beta_0 + b_q + \varepsilon_{q,d},\]
  \[\quad
  b_q\sim \mathcal{N}(0,\tau^2),\ \varepsilon_{q,d}\sim \mathcal{N}(0,\sigma^2),
\]
treating each query as a random intercept to account for repeated measures. We then perform a Wald $z$-test of:
$H_0: \beta_0=-0.011$ vs. $H_A: \beta_0 \neq -0.011$. Table~\ref{tab:minskew_wald_summary} summarizes the estimated intercepts, standard errors, Wald $z$‐statistics, and confidence intervals across different values of $k$. 
In all cases, the Wald tests strongly reject the null hypothesis ($p < 0.001$), indicating that the population‐mean deviation is significantly below the reported value for each cutoff.

\begin{table}[h]
\centering

\begin{tabular}{cccccc}
\toprule
$k$ & $\hat{\beta}_0$ & SE & $z$ & $p$ & 95\% CI \\
\midrule
25  & $-0.360$ & 0.030 & $-12.2$ & $< 0.001$ & $[-0.418,\; -0.302]$ \\
50  & $-0.278$ & 0.031 & $-9.03$ & $< 0.001$ & $[-0.338,\; -0.217]$ \\
75  & $-0.247$ & 0.032 & $-7.79$ & $< 0.001$ & $[-0.309,\; -0.185]$ \\
100 & $-0.213$ & 0.031 & $-6.85$ & $< 0.001$ & $[-0.273,\; -0.152]$ \\
\bottomrule
\end{tabular}
\caption{Summary of Wald $z$‐tests evaluating whether MinSkew@$k$ values are significantly different from zero across different cutoff ranks $k$.}
\label{tab:minskew_wald_summary}
\end{table}

\subsection{Auditing for Race-based Post-Processing}\label{app:results_race}

LinkedIn has not reported any fairness analysis for racial groups in their original key papers on bias \citep{geyik2019fairness}. However, both the law and their very recent work \citep{badrinarayanan2024privacy} suggests that achieving fairness in rankings with respect to race is needed.
Therefore, we perform the first-ever analysis of disparities across racial groups on LTS. We extend our analysis of gender groups to examine whether similar trends emerge across racial groups.

\begin{figure}[h]
    \centering
    \includegraphics[width=1\columnwidth]{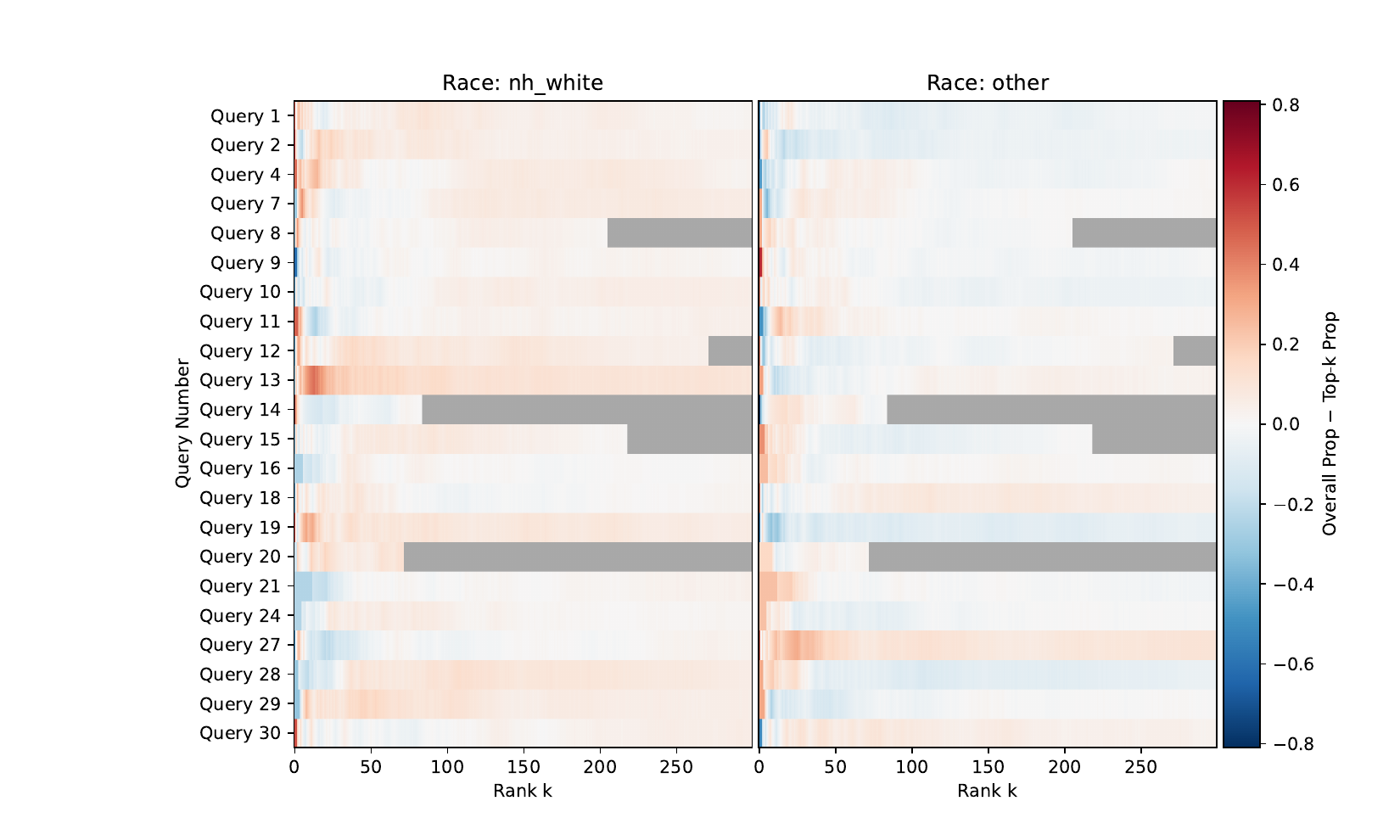}
    \caption{Deviation between the observed top-$k$ race proportions and the overall candidate pool proportions for the set of queries for which we scraped the full list of returned candidates. Each row corresponds to a query, with race-wise deviations shown across rank positions up to $k = 300$. Gray areas indicate ranks beyond the total number of returned candidates for that query (i.e., the candidate pool was smaller than 300). Red values indicate under-representation relative to the overall group proportion, while blue values indicate over-representation.}
    \label{fig:specific_queries_race_deviation}
\end{figure}

\begin{figure}[h]
    \centering
    \includegraphics[width=1\columnwidth]{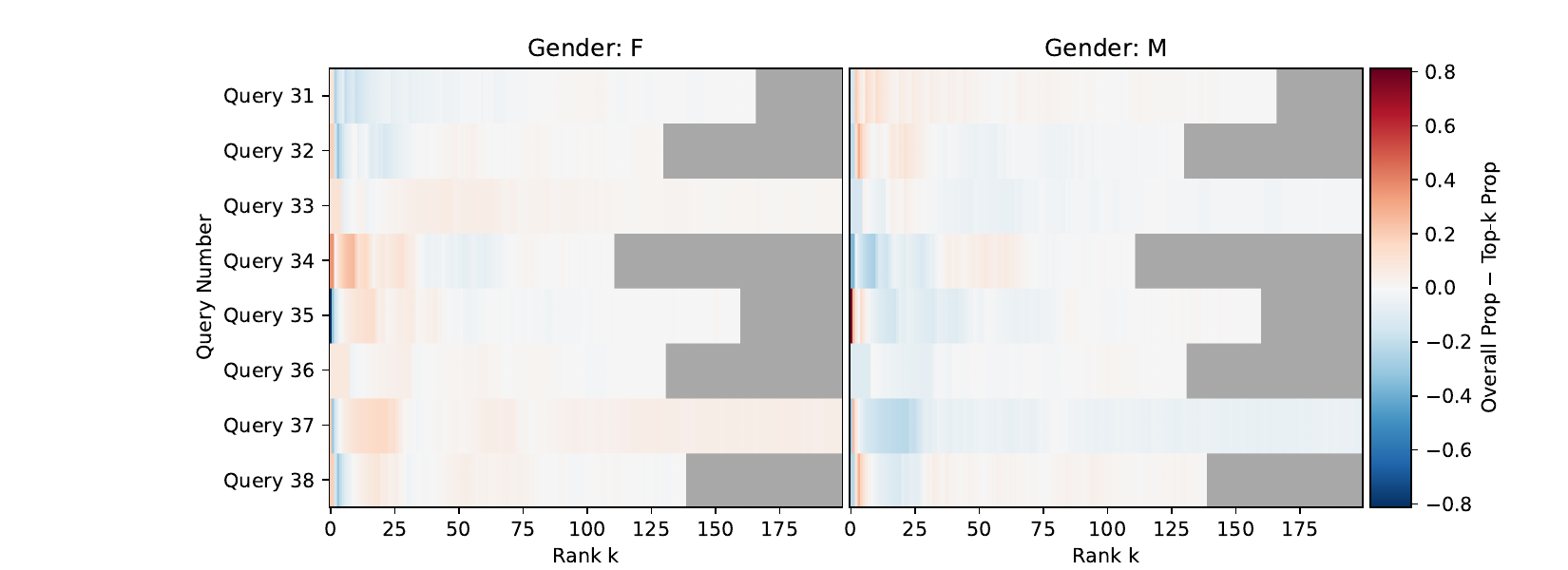}
    \caption{Deviation between the observed top-$k$ race proportions and the overall candidate pool proportions for the set of queries with less than 1\% missing data up to rank 200. Gray areas indicate ranks beyond the total number of returned candidates for that query.}
    \label{fig:specific_queries_race_deviation_no_missing}
\end{figure}

We perform an analogous analysis to that in Section~\ref{subsec:group-deviation}, this time focusing on two racial groups: Non-Hispanic White (nh\_white) and Other. See the discussion in Section~\ref{sec:data_labeling} for the reason for this two-group clustering. We find a pattern of under-representation
at the top ranks, although the under-representation is less consistent across groups.  Given the lower accuracy of our racial inference (Section~\ref{sec:data_labeling}) and the additional noise it introduces, we restrict our analysis to the third query group in which we have a better estimate of $p_i^*$ compared to the general queries. 
Figures~\ref{fig:specific_queries_race_deviation} and~\ref{fig:specific_queries_race_deviation_no_missing} depict deviations from overall group proportions for our third query set, filtered to under 15\% (in $Q_3$) and 1\% ($Q_3^\prime$) missing candidates, respectively. For queries with less than 15\% missing candidates, the White cohort is generally under-represented at higher ranks, though the reverse pattern appears at the very top positions. When we tighten the filter to queries with less than 1\% missing members, deviations no longer follow a consistent trend; some queries exhibit over-representation while others show under-representation, both near the top and further down. 

\begin{figure}[h]
    \centering
    \includegraphics[width=1\linewidth]{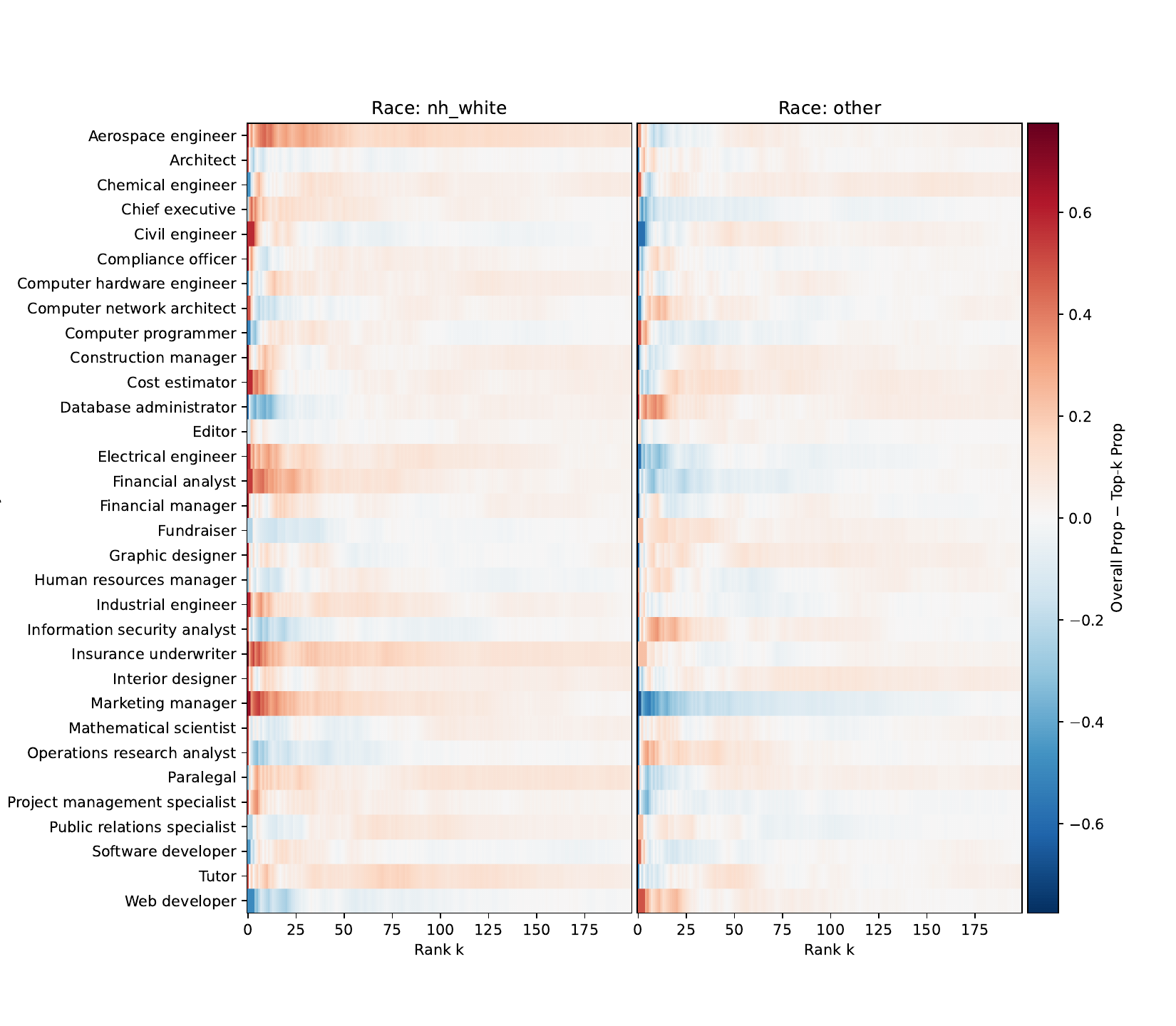}
    \caption{Deviation from overall group proportions at each rank $k$ for two racial categories: Non-Hispanic White (left) and Other (right), across general queries with less than 15\% missing candidates. Red indicates under-representation relative to the group’s overall proportion in the candidate pool, while blue indicates over-representation.}
    \label{fig:race_deviation_actual}
\end{figure}

\begin{figure}[h]
    \centering
    \includegraphics[width=1\linewidth]{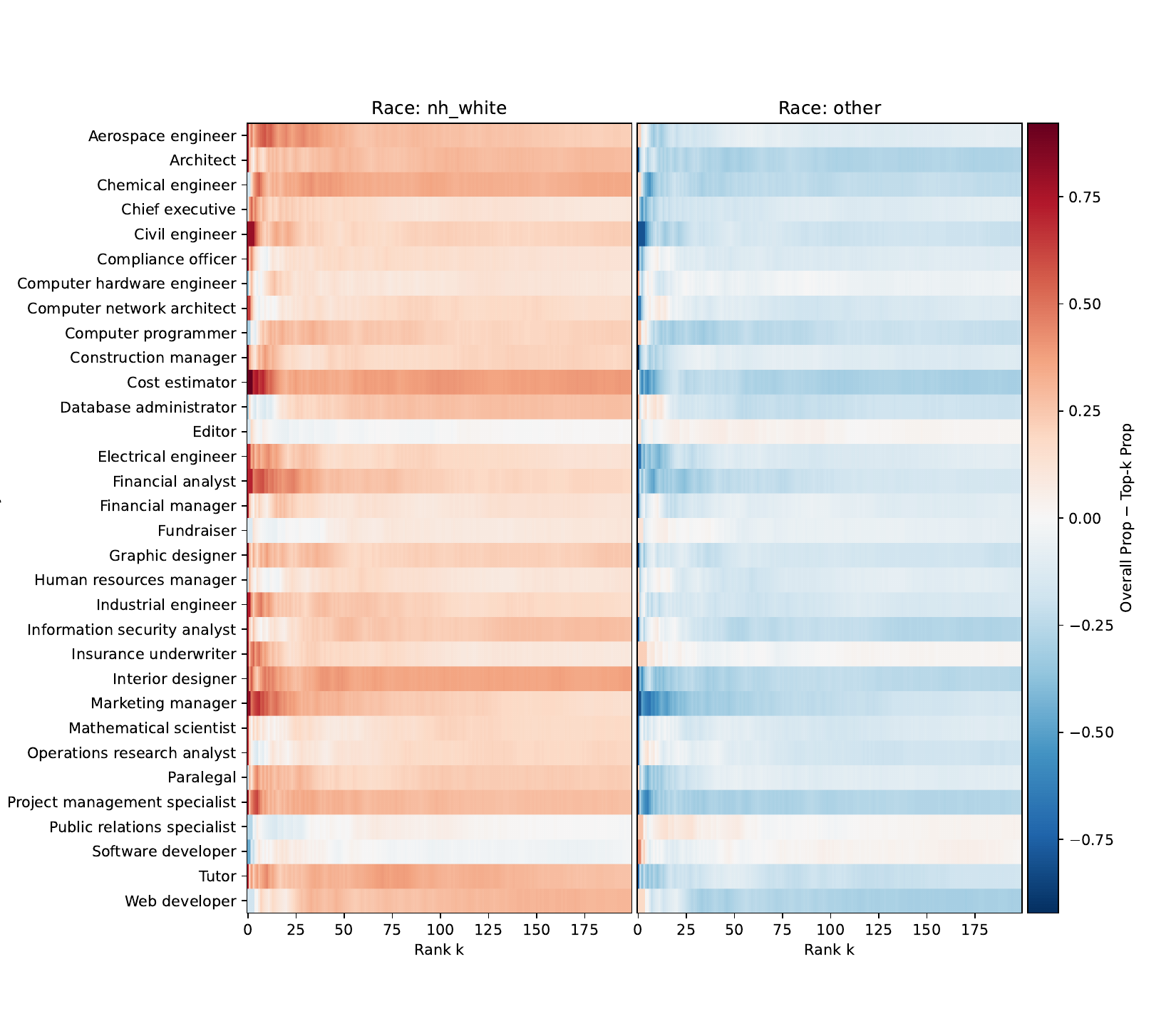}
    \caption{Deviation from BLS proportions at each rank $k$ for two racial categories: Non-Hispanic White (left) and Other (right), across general queries with less than 15\% missing candidates. Red indicates under-representation relative to the group’s overall proportion in the candidate pool, while blue indicates over-representation.}
    \label{fig:race_deviation_BLS}
\end{figure}

 Figure \ref{fig:race_deviation_actual} presents the deviation from overall group proportions within the top 200 ranks for the set of general queries containing less than 15\% missing candidates. Note that, unlike what we observed in our analysis for gender bias, the deviation does not always converge to zero at rank 200, despite being benchmarked against the computed overall proportion up to that rank. This is because the full candidate pool includes a third group, candidates with missing information, that is excluded from the up-to-rank-$k$ calculations for either group. As shown, both racial groups exhibit substantial deviations across different queries, particularly in the top ranks. No single group is uniformly over- or under-represented across all queries. These deviations generally diminish with increasing $k$, suggesting that representation aligns more closely with overall proportions deeper in the ranking. 
 Figure \ref{fig:race_deviation_BLS} shows deviations in ranked group proportions relative to the BLS’s reported racial distribution. Unlike Figure \ref{fig:race_deviation_actual}, here the White cohort is consistently under-represented across nearly all queries, and the magnitude of deviation remains large even at deeper ranks. Given the limitations of both approaches discussed in Section \ref{app:results_gender}, we cannot draw definitive conclusions; however, both analyses do reveal evidence of disparity across racial groups.

Because MinSkew is highly sensitive to noise and our racial‐inference process introduces substantial uncertainty, we consider the metric unreliable for analyzing racial groups in this scenario. The tendency for deviations to diminish as rank increases, across both filters, may indicate that LinkedIn applies race-aware post-processing to promote equal treatment.
Alternatively, these results may reflect underlying similarities in the relevance score distributions across the racial groups we audited, suggesting that the observed patterns could be a byproduct of those similarities.

\clearpage
\subsection{Temporal Aspects of Fairness} \label{app:results_temporal}
We conduct a temporal analysis similar to that in Section~\ref{sec:temporal_analysis}, applying it to our set of general queries ($Q_2$), and extend it to racial groups.
For gender, we observe patterns consistent with those seen earlier --- such as higher churn rates for certain groups --- while for race, no consistent trends emerge across query sets. To assess whether group size influences churn, we examine correlations between group proportions and churn rates and find no meaningful relationship. 
Finally, statistical tests confirm that the observed differences between groups are significant, reinforcing the presence of temporal disparities in exposure. We then look for correlation in group size and churn rates that might explain disparities, and find no negative or positive correlation. 
Finally, the results of our statistical tests show that the observed between-group disparities are significant enough.

\begin{figure}[h]
    \centering
    \includegraphics[width=1\linewidth]{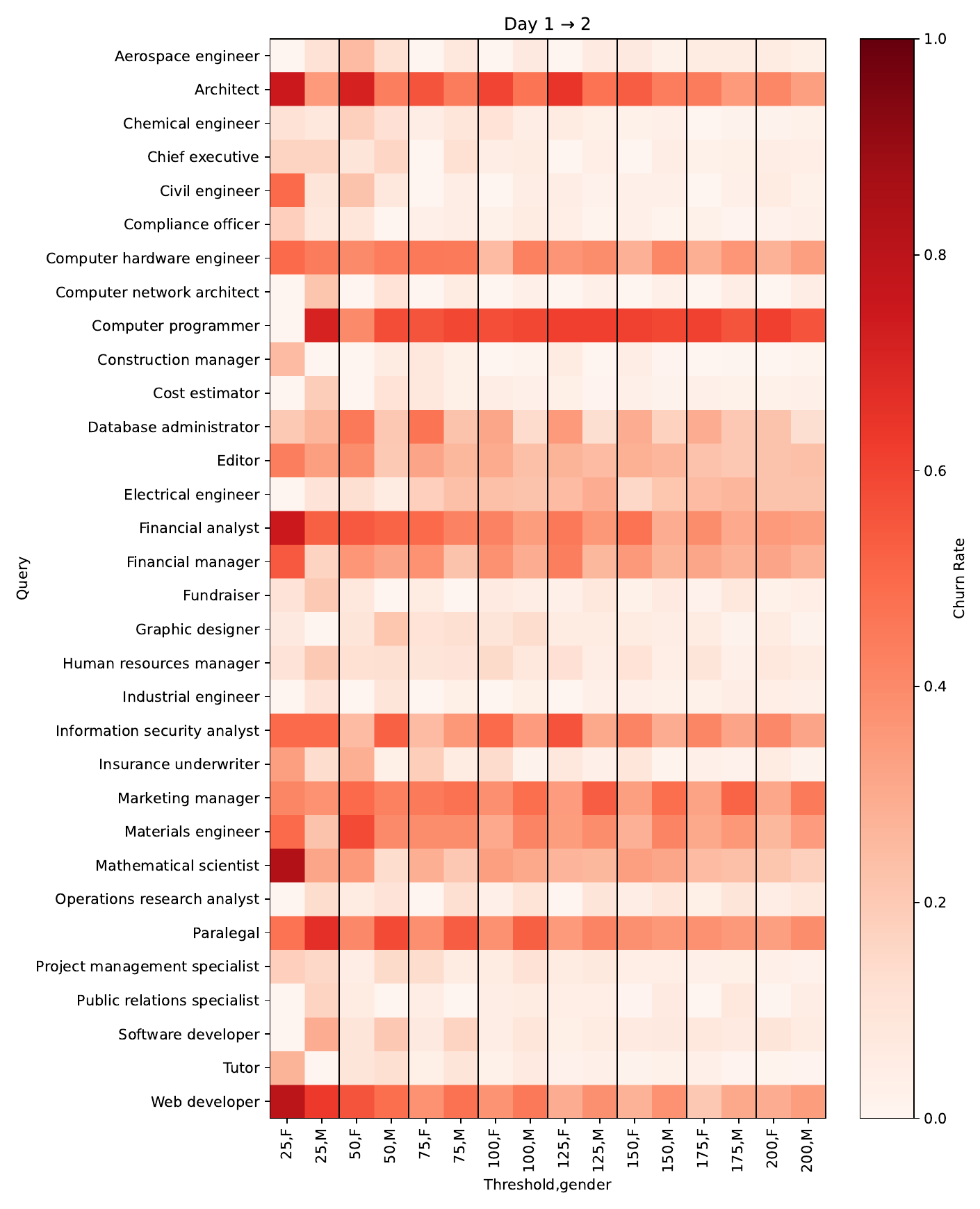}
    \caption{Churn rates by gender across our set of general queries at 25-rank intervals from the first day to the second day.}
    \label{fig:churn_rate_general_1to2}
\end{figure}

\begin{figure}[h]
    \centering
    \includegraphics[width=1\linewidth]{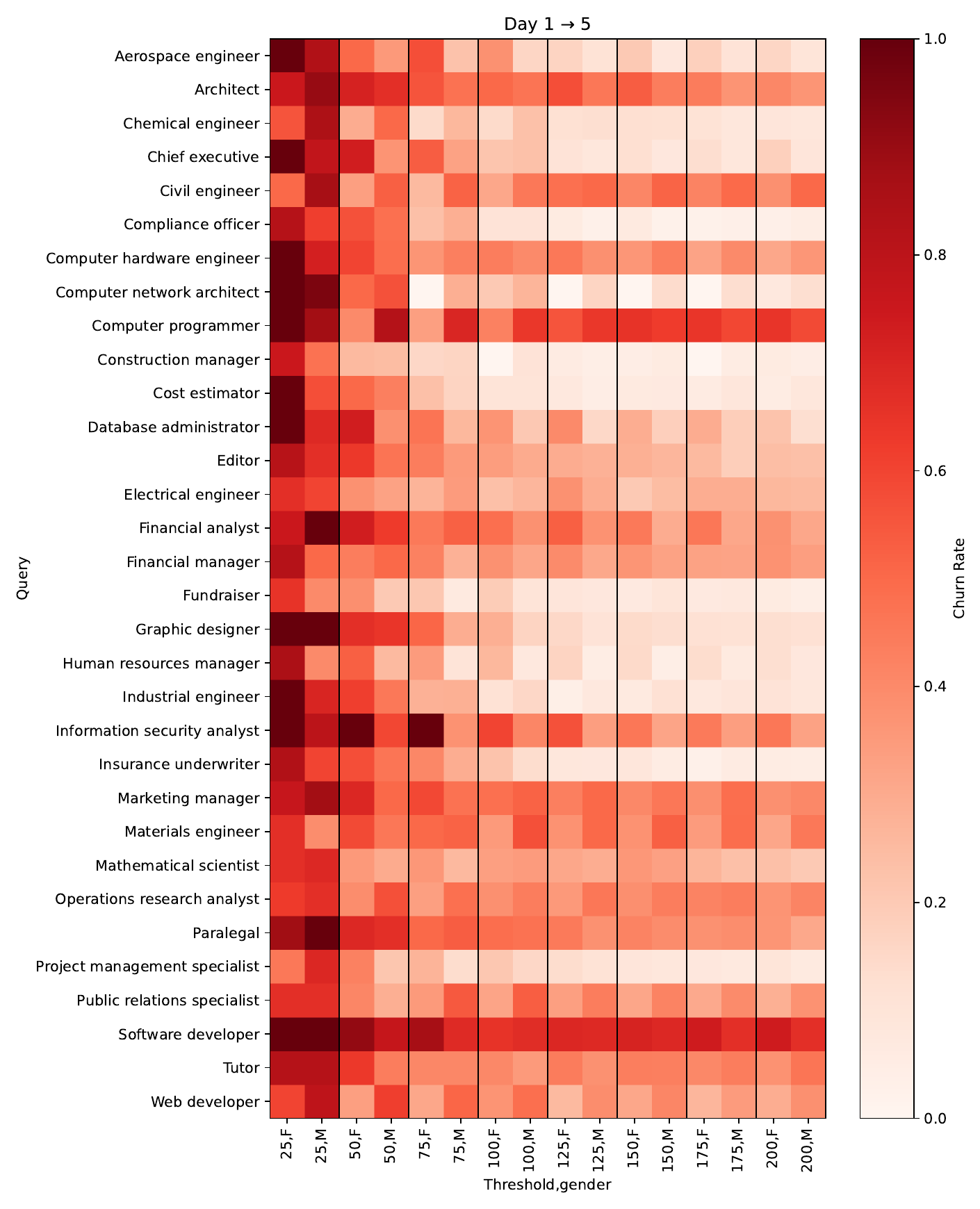}
    \caption{Churn rates by gender across our set of general queries at 25-rank intervals from the first day to the fifth day.}
    \label{fig:churn_rate_general_1to5}
\end{figure}

\begin{figure}[h]
    \centering
    \includegraphics[width=1\linewidth]{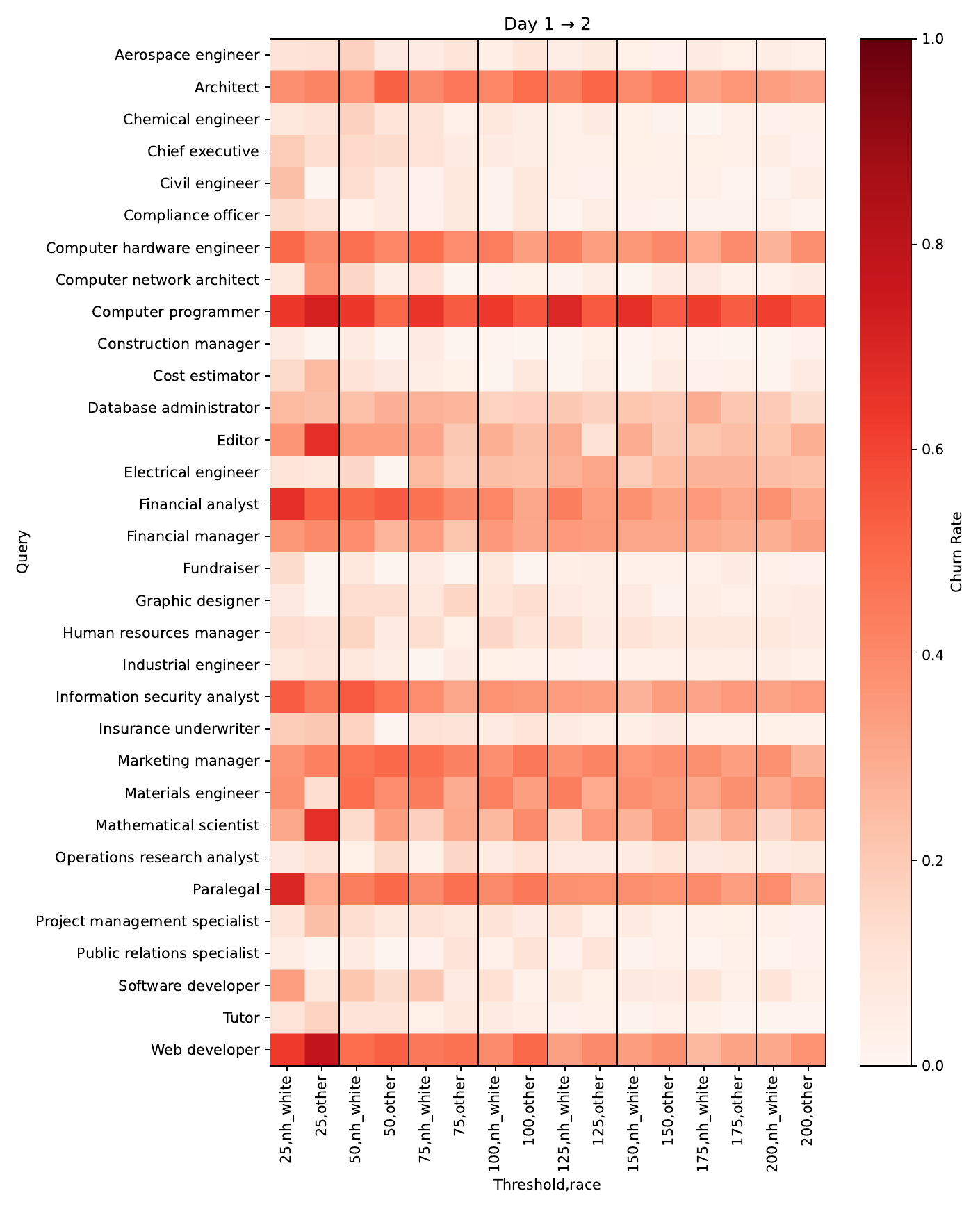}
    \caption{Churn rates by race across our set of general queries at 25-rank intervals from the first day to the second day.}
    \label{fig:churn_rate_race_general_1to2}
\end{figure}

\begin{figure}[h]
    \centering
    \includegraphics[width=1\linewidth]{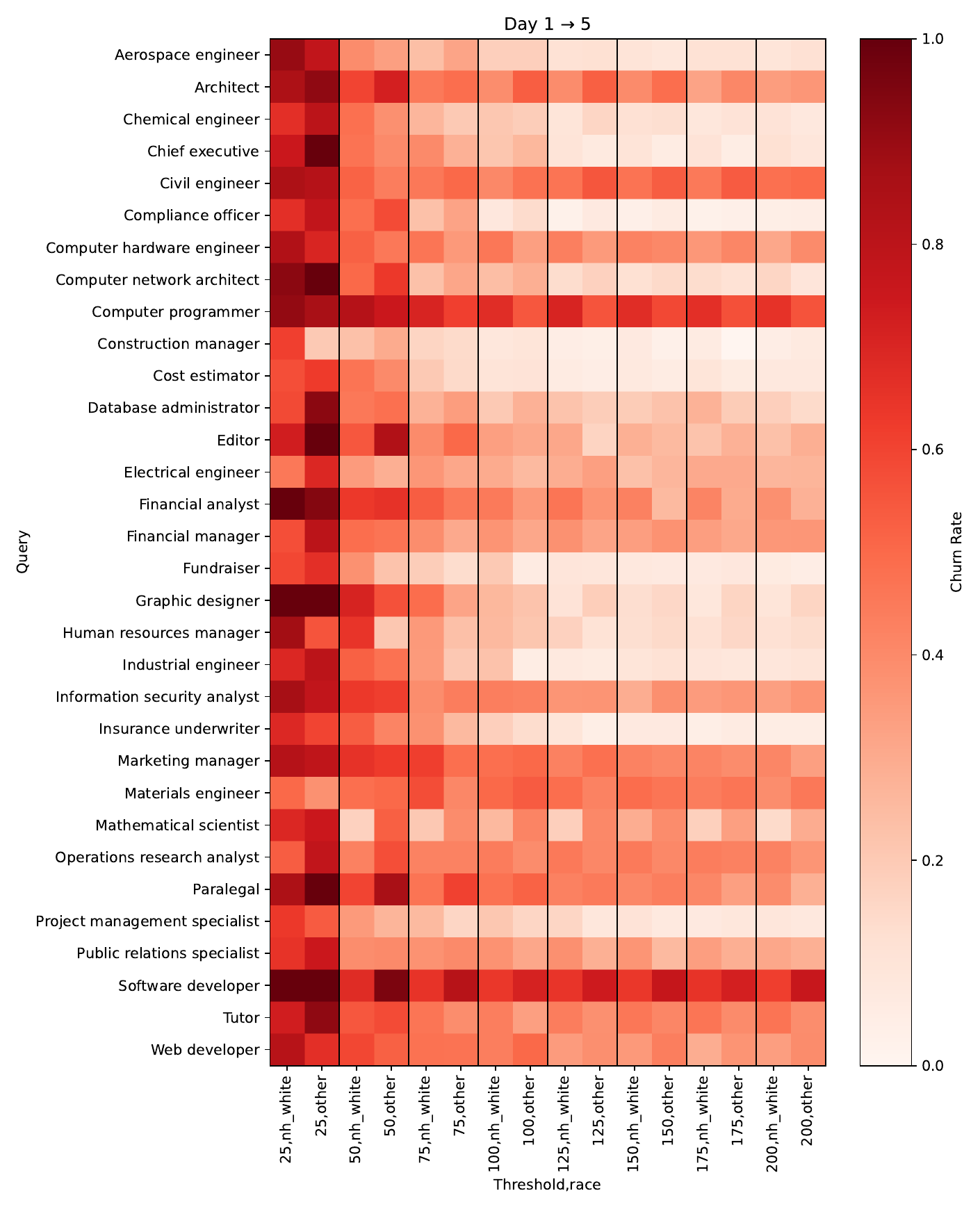}
    \caption{Churn rates by race across our set of general queries at 25-rank intervals from the first day to the fifth day.}
    \label{fig:churn_rate_race_general_1to5}
\end{figure}

Figures~\ref{fig:churn_rate_general_1to2} and \ref{fig:churn_rate_general_1to5} show $\text{Churn}_{g_i}^{1 \rightarrow 2}(k)$ and $\text{Churn}_{g_i}^{1 \rightarrow 5}(k)$ for $k \in \{25, 50, \ldots, 200\}$ and $g_i \in \{F, M\}$ for the set of general queries. Figures~\ref{fig:churn_rate_race_general_1to2} and \ref{fig:churn_rate_race_general_1to5} show the corresponding churn rates for the Non-Hispanic White and Other racial groups.
Even when gender groups exhibit similar churn from day 1 to day 2, by day 5, the gap widens substantially, with women consistently experiencing higher turnover. For racial groups, churn rates similarly intensify over time, but the magnitude of the gap between churn rates from days 2 to 5 varies across queries.

\begin{figure}[h!]
    \centering
    \includegraphics[width=0.49\linewidth]{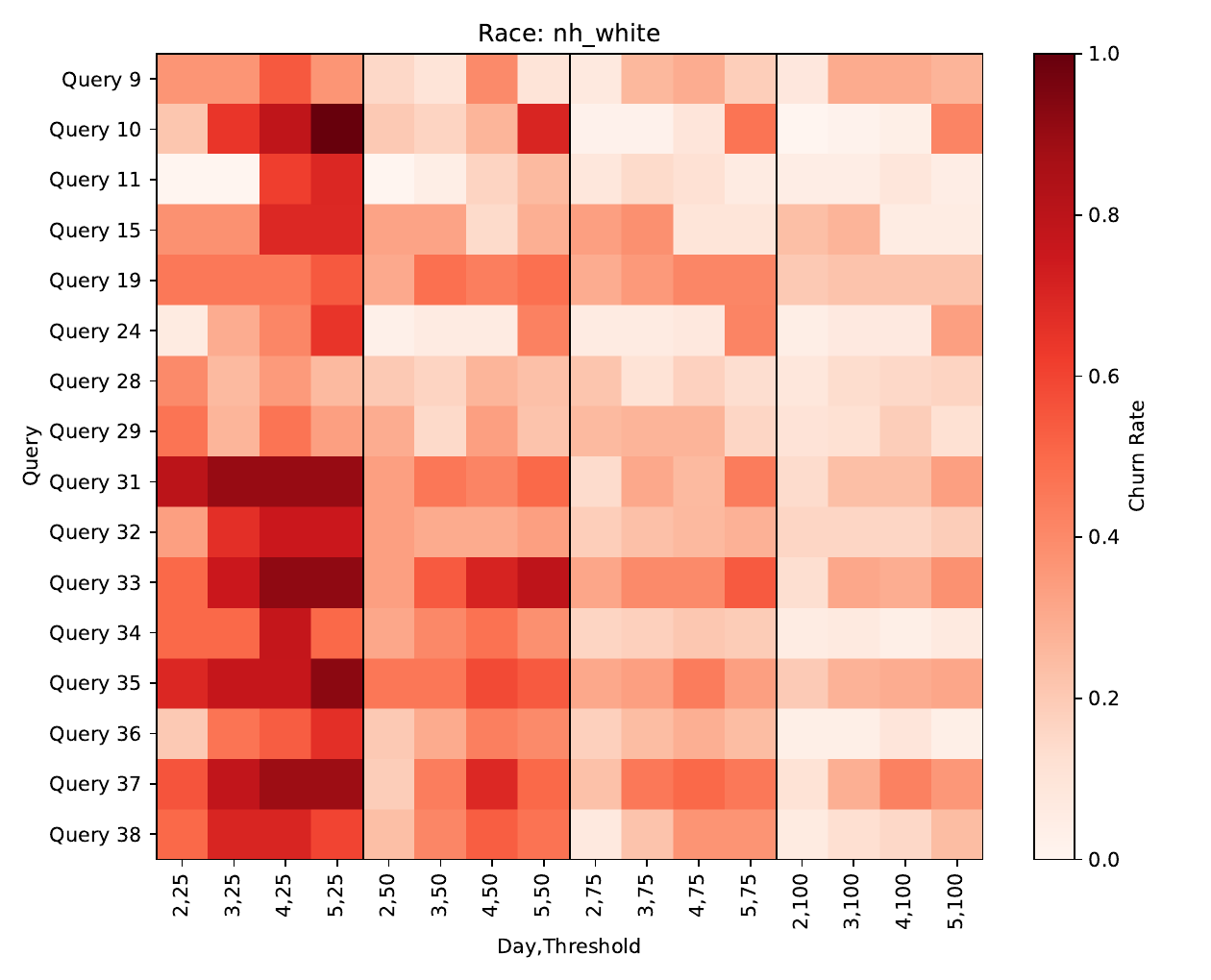}
    \includegraphics[width=0.49\linewidth]{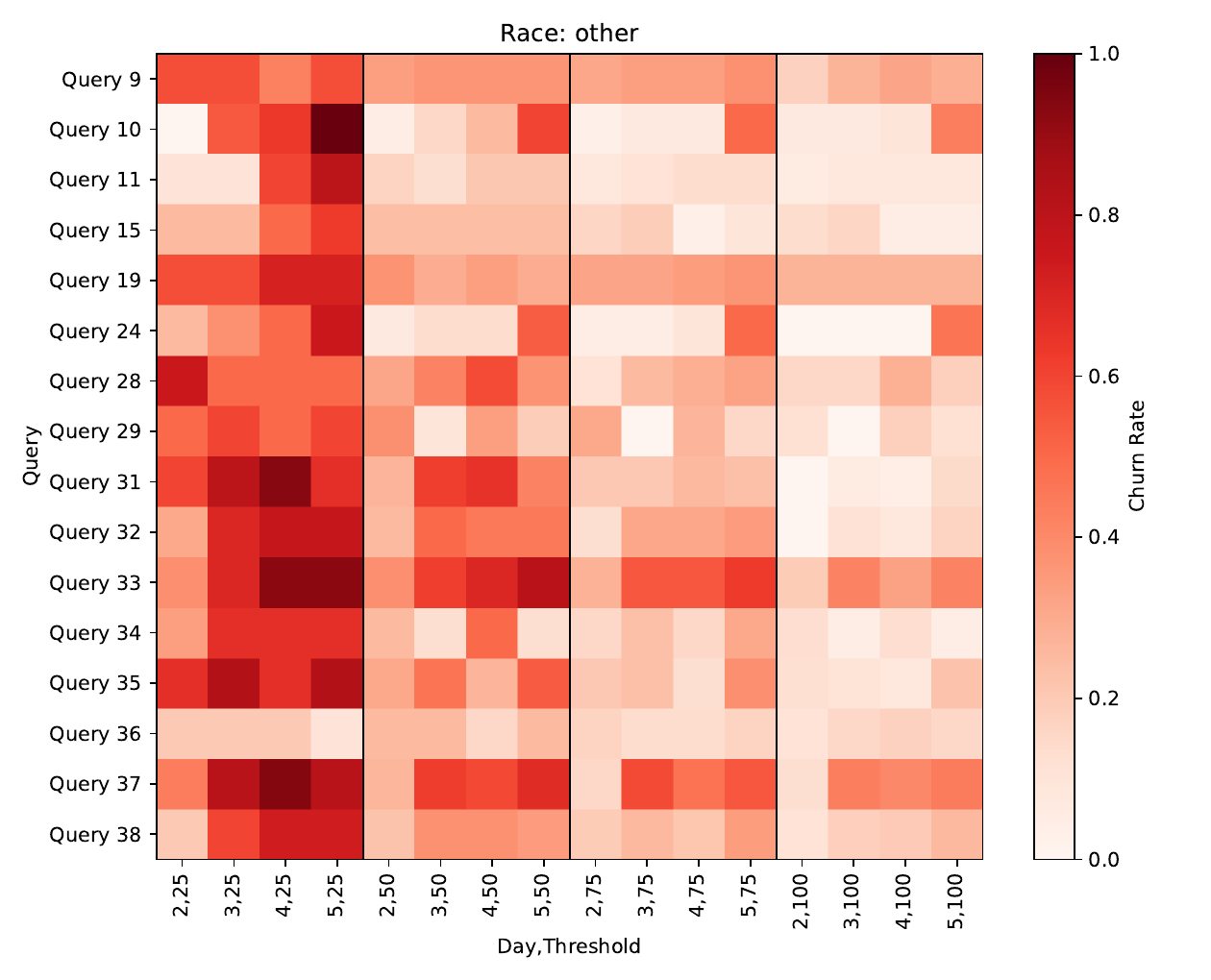}
    \caption{(\textbf{Left}): Heatmap of day‐to‐day churn rates \(\mathrm{Churn}_{g_i}^{1\rightarrow i}(k)\) for the non-Hispanic White candidates across position‐specific queries (with less than 15\% missing candidates) over five consecutive days, evaluated at top‐\(k\) cutoffs \(k\in\{25,50,\dots,100\}\). (\textbf{Right}): Same plot for the ``other'' racial group.}
    \label{fig:churn_rate_specific_white}
\end{figure}

Figure~\ref{fig:churn_rate_specific_white} shows $\text{Churn}_{g_i}^{1 \rightarrow i}(k)$ for $k \in \{25, 50, \ldots, 200\}$ and $i \in \{2,3,4,5\}$ for non-hispanic Whites and ``Other'' group, in the set of position-specific queries with less than 15\% missing candidates and 5 consecutive days of data ($Q_3 \cup Q'_3$). Although churn rates vary in both magnitude and temporal patterns, these variations are not consistent across queries for these two groups, and no clear overall pattern is observed when comparing the two groups.

\begin{figure}[h!]
    \centering
    \includegraphics[width=0.49\linewidth]{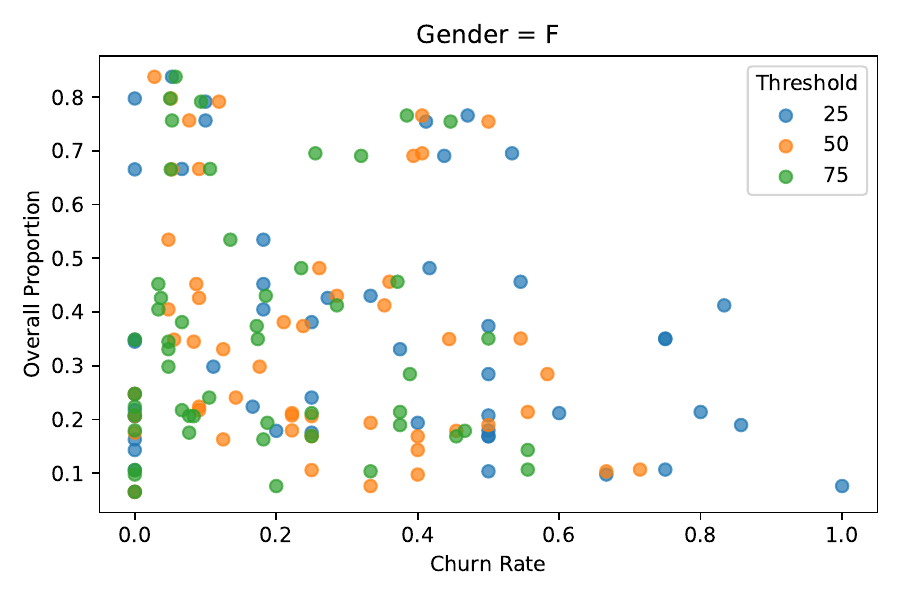}
    \includegraphics[width=0.49\linewidth]{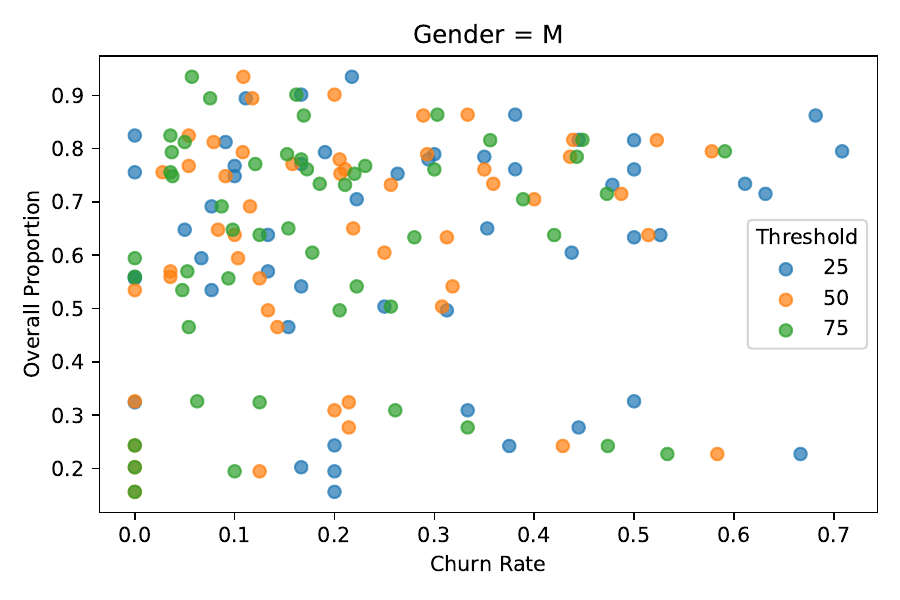}
    \caption{(\textbf{Left}): Scatter plot of day 1$\rightarrow$ 2 churn rate versus overall group proportion for the female group. Points are colored by rank cutoff \(k\in\{25,50,75\}\). (\textbf{Right}): Same plot for the male group.}
    \label{fig:churn_vs_prop_F}
\end{figure}

When interpreting differences in churn rates across groups, we must rule out confounding by group size. To confirm that women’s higher churn is not merely a consequence of their lower overall representation, we plot the day 1$\rightarrow$2 churn rate against each group’s overall proportion for queries with less than 15\% missing data and at least 75 candidates at thresholds 25, 50, and 75, indicated with different colors. 
Figure~\ref {fig:churn_vs_prop_F} reveals no clear positive or negative correlation in either group.

\subsubsection{Statistical Analysis} \label{app:statistical_test_temporal}
To evaluate whether men and women differ significantly in churn rates across different days, we fit a series of linear mixed-effects models of the form:
\[
\text{churn}_{q,d}
= \beta_{0}
+ \beta_{\mathrm{g}}\;\mathbb{I}(\mathrm{gender}_{q,d} = \mathrm{M})
+ \beta_{\mathrm{d}}\;\mathrm{d}
+ b_{q}
+ \varepsilon_{q,d},
\]
\[
\text{where}\quad
b_{q}\sim \mathcal{N}(0,\tau^{2}),\quad
\varepsilon_{q,d}\sim \mathcal{N}(0,\sigma^{2}),
\]
To assess whether gender-based differences in churn persist even after accounting for time progression, we fit a mixed-effects model predicting churn from day 1 to subsequent days (2 through 5), with day and gender as fixed effects. The model includes 384 observations across 48 queries that had at least 100 candidates and fewer than 15\% missing. 

For all queries $q$ with maximum ranks above 100 and a missing candidate rate of less than 15\% from day 1 to day 2, 3, 4, and 5. This gives us 384 observations across 48 queries. We tested the hypothesis that, after accounting for the fact that churn systematically increases over time, there remains a persistent difference in churn rates between men and women. 
$H_0: \beta_M = 0$ versus $H_A: \beta_M \neq 0$.  As shown in Table~\ref{tab:gender-churn-model}, we observe statistically significant negative coefficients for the male group at \(k=25\) and \(k=50\), indicating that male candidates churn significantly less than female candidates at early ranks, even after controlling for the increasing trend in churn over time. These disparities dissipate at \(k=75\) and \(k=100\), suggesting that gender-based differences in candidate persistence are concentrated at the top of the rankings. A similar model fitted for racial groups revealed no significant churn disparities across any $k$.

\begin{table}[ht]
\centering
\begin{tabular}{ccccccc}
\toprule
$k$ & $\hat{\beta}_{\mathrm{M}}$ & SE & $z$ & $p$ (group) & $\hat{\beta}_{\mathrm{day}}$ & $p$ (day) \\
\midrule
25  & $-0.071$ & $0.021$ & $-3.38$ & $0.001$ & $0.140$ & $<0.001$ \\
50  & $-0.037$ & $0.013$ & $-2.76$ & $0.006$ & $0.075$ & $<0.001$ \\
75  & $0.002$      & $0.023$     & $0.14$     & $0.889$     & $0.051$     & $<0.001$ \\
100 & $0.005$      & $0.009$     & $0.553$     & $0.580$     & $0.037$     & $<0.001$ \\
\bottomrule
\end{tabular}
\caption{Mixed‐effects model results predicting churn ($\text{Churn}_{g_i}^{t_1 \rightarrow t_d}(k)$) across days 2–5 with group ($g_i=$ Male) and day as fixed effects. Negative $\hat{\beta}_{\mathrm{M}}$ values indicate lower churn for male candidates relative to female.}
\label{tab:gender-churn-model}
\end{table}

\subsection{Statistical Summaries of Results}
In this section, we present the average values of all evaluated metrics (deviation at top-$k$, $\text{skew}@k$, and churn rate) computed for each demographic group and query set. For comparison, we also report the corresponding values obtained from a random baseline ranking for each query. 
The random baseline was calculated by assigning random ranks to candidates over all sets of queries.

Table \ref{tab:deviation_summary} presents the average deviation values from top-$k$ rankings for $k \in \{25, 50, 75, 100\}$. 
For the gender attribute, the trends are consistent across both query sets ($Q_2$ and $Q_3 \cup Q_3'$): female groups exhibit higher deviation from overall group proportions, while the differences between groups, as well as the overall deviation, gradually decrease as $k$ increases.
For the race attribute, although deviation generally decreases with larger $k$ values, the two query sets show different patterns in which subgroup's results contain higher average deviation.
We find that while some deviation remains under random orderings, the group-level disparities are markedly smaller than those in the actual rankings, particularly at the top ranks, suggesting that the observed differences are unlikely to be fully explained by random variation alone.

\begin{table}[ht]
\centering
\begin{tabular}{lcccc}
\toprule
\multicolumn{5}{c}{\textbf{$Q_2$}} \\
\midrule
Group & $k=25$ & $k=50$ & $k=75$ & $k=100$ \\
\midrule
F & 0.043 $\pm$ 0.085 & 0.030 $\pm$ 0.064 & 0.028 $\pm$ 0.045 & 0.026 $\pm$ 0.036 \\
M & 0.012 $\pm$ 0.093 & 0.028 $\pm$ 0.076 & 0.028 $\pm$ 0.060 & 0.029 $\pm$ 0.047 \\
nh\_white & 0.035 $\pm$ 0.098 & 0.041 $\pm$ 0.067 & 0.044 $\pm$ 0.055 & 0.039 $\pm$ 0.045 \\
other & 0.025 $\pm$ 0.087 & 0.021 $\pm$ 0.055 & 0.015 $\pm$ 0.043 & 0.018 $\pm$ 0.033 \\
\midrule
\multicolumn{5}{c}{\textbf{$Q_3 \cup Q'_3$}} \\
\midrule
Group & $k=25$ & $k=50$ & $k=75$ & $k=100$ \\
\midrule
F & 0.046 $\pm$ 0.090 & 0.043 $\pm$ 0.093 & 0.048 $\pm$ 0.092 & 0.053 $\pm$ 0.087 \\
M & 0.011 $\pm$ 0.081 & 0.021 $\pm$ 0.068 & 0.021 $\pm$ 0.057 & 0.022 $\pm$ 0.050 \\
nh\_white & 0.019 $\pm$ 0.103 & 0.057 $\pm$ 0.091 & 0.066 $\pm$ 0.085 & 0.074 $\pm$ 0.093 \\
other & 0.041 $\pm$ 0.084 & 0.012 $\pm$ 0.064 & 0.006 $\pm$ 0.065 & 0.002 $\pm$ 0.061 \\
\midrule
\multicolumn{5}{c}{\textbf{Random Baseline $Q_2\cup Q_3 \cup Q_3'$}} \\
\midrule
Group & $k=25$ & $k=50$ & $k=75$ & $k=100$ \\
\midrule
F & 0.027 $\pm$ 0.091 & 0.031 $\pm$ 0.074 & 0.030 $\pm$ 0.070 & 0.032 $\pm$ 0.067 \\
M & 0.031 $\pm$ 0.093 & 0.032 $\pm$ 0.063 & 0.031 $\pm$ 0.056 & 0.029 $\pm$ 0.048 \\
nh\_white & 0.038 $\pm$ 0.100 & 0.041 $\pm$ 0.078 & 0.037 $\pm$ 0.065 & 0.038 $\pm$ 0.060 \\
other & 0.021 $\pm$ 0.087 & 0.023 $\pm$ 0.066 & 0.026 $\pm$ 0.056 & 0.025 $\pm$ 0.046 \\
\bottomrule
\end{tabular}
\caption{Deviation from $k$ statistical summaries across groups for different query sets.}
\label{tab:deviation_summary}
\end{table}

Table~\ref{tab:skewk_summary} reports the average $\text{skew}@k$ values for $k \in \{25, 50, 75, 100\}$ across demographic groups, query sets, and the random baseline.
Notably, all average values correspond to an overall MinSkew@$k$ that is lower than the $-0.011$ reported by \citet{geyik2019fairness}.
For gender, both query sets show an upward trend in $\text{skew}@k$ (indicating improved balance) as $k$ increases, along with a decreasing gap between groups.
For race, the trends are less consistent across the two query sets, and there is no clear pattern regarding which group exhibits lower average skew, but in both cases, the metric improves with larger $k$.
As with deviation, the random baseline shows smaller differences between groups for both gender and race, suggesting that the disparities observed in the actual rankings are not due to random variation alone.

\begin{table}[ht]
\centering
\begin{tabular}{lcccc}
\toprule
\multicolumn{5}{c}{\textbf{$Q_2$}} \\
\midrule
Group & $k=25$ & $k=50$ & $k=75$ & $k=100$ \\
\midrule
F & -0.253 $\pm$ 0.459 & -0.111 $\pm$ 0.230 & -0.109 $\pm$ 0.213 & -0.104 $\pm$ 0.136 \\
M & 0.012 $\pm$ 0.153 & -0.060 $\pm$ 0.169 & -0.044 $\pm$ 0.130 & -0.046 $\pm$ 0.111 \\
nh\_white & -0.088 $\pm$ 0.208 & -0.071 $\pm$ 0.112 & -0.068 $\pm$ 0.103 & -0.066 $\pm$ 0.083 \\
other & -0.055 $\pm$ 0.248 & -0.059 $\pm$ 0.161 & -0.037 $\pm$ 0.125 & -0.038 $\pm$ 0.105 \\
\midrule
\multicolumn{5}{c}{\textbf{$Q_3 \cup Q_3'$}} \\
\midrule
Group & $k=25$ & $k=50$ & $k=75$ & $k=100$ \\
\midrule
F & -0.144 $\pm$ 0.391 & -0.095 $\pm$ 0.198 & -0.093 $\pm$ 0.116 & -0.090 $\pm$ 0.114 \\
M & 0.040 $\pm$ 0.086 & -0.006 $\pm$ 0.073 & 0.006 $\pm$ 0.057 & 0.002 $\pm$ 0.049 \\
nh\_white & -0.020 $\pm$ 0.182 & -0.026 $\pm$ 0.109 & -0.033 $\pm$ 0.099 & -0.043 $\pm$ 0.085 \\
other & -0.034 $\pm$ 0.316 & 0.008 $\pm$ 0.134 & 0.032 $\pm$ 0.105 & 0.028 $\pm$ 0.092 \\
\midrule
\multicolumn{5}{c}{\textbf{Random Baseline $Q_2\cup Q_3 \cup Q_3'$}} \\
\midrule
Group & $k=25$ & $k=50$ & $k=75$ & $k=100$ \\
\midrule
F & -0.105 $\pm$ 0.338 & -0.089 $\pm$ 0.244 & -0.100 $\pm$ 0.208 & -0.072 $\pm$ 0.142 \\
M & -0.064 $\pm$ 0.235 & -0.063 $\pm$ 0.185 & -0.043 $\pm$ 0.138 & -0.037 $\pm$ 0.097 \\
nh\_white & -0.066 $\pm$ 0.182 & -0.065 $\pm$ 0.103 & -0.052 $\pm$ 0.087 & -0.044 $\pm$ 0.080 \\
other & -0.087 $\pm$ 0.289 & -0.037 $\pm$ 0.175 & -0.047 $\pm$ 0.145 & -0.052 $\pm$ 0.121 \\
\bottomrule
\end{tabular}
\caption{Statistical summaries for ($\text{skew}@k$) across groups for different query sets. The random baseline was calculated by assigning random ranks to candidates over all sets of queries.}
\label{tab:skewk_summary}
\end{table}

Tables~\ref{tab:Q2_churn_summary}, \ref{tab:q3_churn_cummary}, and \ref{tab:random_churn_summary} present the average churn rates across queries in $Q_2$, $Q_3 \cup Q_3'$, and a random variation of the rankings for queries in $Q_2 \cup Q_3 \cup Q_3'$, respectively.
Each table reports averages computed over day pairs separated by the same temporal distance: for example, the ``1 day apart'' subtables aggregate churn across all consecutive day pairs ($1 \to 2, 2 \to 3, 3 \to 4, 4 \to 5$), the ``2 days apart'' subtables average over pairs ($1 \to 3, 2 \to 4, 3 \to 5$), and so on. Across all query sets, churn rates increase with greater temporal distance between rankings. The female group shows higher average churn overall, whereas the trends for racial groups vary and do not display a consistent pattern.

\begin{table}[ht]
\centering
\begin{tabular}{lcccc}
\toprule
\multicolumn{5}{c}{\textbf{1 day apart}} \\
\midrule
Group & $k=25$ & $k=50$ & $k=75$ & $k=100$ \\
\midrule
F & 0.393 $\pm$ 0.288 & 0.254 $\pm$ 0.201 & 0.161 $\pm$ 0.164 & 0.127 $\pm$ 0.138 \\
M & 0.344 $\pm$ 0.200 & 0.206 $\pm$ 0.150 & 0.153 $\pm$ 0.137 & 0.120 $\pm$ 0.138 \\
nh\_white & 0.350 $\pm$ 0.196 & 0.219 $\pm$ 0.146 & 0.160 $\pm$ 0.135 & 0.126 $\pm$ 0.127 \\
other & 0.360 $\pm$ 0.231 & 0.209 $\pm$ 0.159 & 0.145 $\pm$ 0.130 & 0.117 $\pm$ 0.136 \\
\midrule
\multicolumn{5}{c}{\textbf{2 days apart}} \\
\midrule
Group & $k=25$ & $k=50$ & $k=75$ & $k=100$ \\
\midrule
F & 0.651 $\pm$ 0.277 & 0.377 $\pm$ 0.189 & 0.243 $\pm$ 0.189 & 0.191 $\pm$ 0.164 \\
M & 0.587 $\pm$ 0.206 & 0.310 $\pm$ 0.157 & 0.226 $\pm$ 0.161 & 0.181 $\pm$ 0.168 \\
nh\_white & 0.592 $\pm$ 0.197 & 0.325 $\pm$ 0.157 & 0.236 $\pm$ 0.147 & 0.190 $\pm$ 0.151 \\
other & 0.614 $\pm$ 0.234 & 0.329 $\pm$ 0.168 & 0.222 $\pm$ 0.156 & 0.177 $\pm$ 0.163 \\
\midrule
\multicolumn{5}{c}{\textbf{3 days apart}} \\
\midrule
Group & $k=25$ & $k=50$ & $k=75$ & $k=100$ \\
\midrule
F & 0.774 $\pm$ 0.234 & 0.474 $\pm$ 0.180 & 0.316 $\pm$ 0.187 & 0.251 $\pm$ 0.161 \\
M & 0.680 $\pm$ 0.191 & 0.392 $\pm$ 0.154 & 0.290 $\pm$ 0.162 & 0.242 $\pm$ 0.174 \\
nh\_white & 0.696 $\pm$ 0.179 & 0.418 $\pm$ 0.149 & 0.305 $\pm$ 0.144 & 0.251 $\pm$ 0.152 \\
other & 0.725 $\pm$ 0.198 & 0.420 $\pm$ 0.172 & 0.289 $\pm$ 0.161 & 0.238 $\pm$ 0.169 \\
\midrule
\multicolumn{5}{c}{\textbf{4 days apart}} \\
\midrule
Group & $k=25$ & $k=50$ & $k=75$ & $k=100$ \\
\midrule
F & 0.809 $\pm$ 0.170 & 0.546 $\pm$ 0.178 & 0.396 $\pm$ 0.197 & 0.321 $\pm$ 0.150 \\
M & 0.730 $\pm$ 0.181 & 0.474 $\pm$ 0.157 & 0.360 $\pm$ 0.160 & 0.319 $\pm$ 0.172 \\
nh\_white & 0.743 $\pm$ 0.152 & 0.508 $\pm$ 0.136 & 0.387 $\pm$ 0.140 & 0.328 $\pm$ 0.151 \\
other & 0.775 $\pm$ 0.187 & 0.513 $\pm$ 0.182 & 0.366 $\pm$ 0.152 & 0.316 $\pm$ 0.166 \\
\bottomrule
\end{tabular}
\caption{Mean churn rates for each group, averaged across all queries and all rankings separated by the same number of days (e.g., one day apart, two days apart, etc.) within $Q_2$.}
\label{tab:Q2_churn_summary}
\end{table}

\begin{table}[ht]
\centering
\begin{tabular}{lcccc}
\toprule
\multicolumn{5}{c}{\textbf{1 day apart}} \\
\midrule
Group & $k=25$ & $k=50$ & $k=75$ & $k=100$ \\
\midrule
F & 0.446 $\pm$ 0.279 & 0.262 $\pm$ 0.218 & 0.200 $\pm$ 0.153 & 0.114 $\pm$ 0.113 \\
M & 0.376 $\pm$ 0.198 & 0.263 $\pm$ 0.145 & 0.185 $\pm$ 0.121 & 0.120 $\pm$ 0.097 \\
nh\_white & 0.398 $\pm$ 0.206 & 0.261 $\pm$ 0.154 & 0.193 $\pm$ 0.128 & 0.115 $\pm$ 0.096 \\
other & 0.380 $\pm$ 0.214 & 0.270 $\pm$ 0.151 & 0.196 $\pm$ 0.134 & 0.126 $\pm$ 0.107 \\
\midrule
\multicolumn{5}{c}{\textbf{2 days apart}} \\
\midrule
Group & $k=25$ & $k=50$ & $k=75$ & $k=100$ \\
\midrule
F & 0.556 $\pm$ 0.326 & 0.320 $\pm$ 0.230 & 0.228 $\pm$ 0.161 & 0.138 $\pm$ 0.122 \\
M & 0.531 $\pm$ 0.229 & 0.306 $\pm$ 0.179 & 0.225 $\pm$ 0.144 & 0.151 $\pm$ 0.122 \\
nh\_white & 0.531 $\pm$ 0.230 & 0.309 $\pm$ 0.187 & 0.232 $\pm$ 0.148 & 0.153 $\pm$ 0.118 \\
other & 0.550 $\pm$ 0.232 & 0.321 $\pm$ 0.200 & 0.234 $\pm$ 0.156 & 0.153 $\pm$ 0.127 \\
\midrule
\multicolumn{5}{c}{\textbf{3 days apart}} \\
\midrule
Group & $k=25$ & $k=50$ & $k=75$ & $k=100$ \\
\midrule
F & 0.687 $\pm$ 0.226 & 0.362 $\pm$ 0.183 & 0.253 $\pm$ 0.160 & 0.176 $\pm$ 0.125 \\
M & 0.634 $\pm$ 0.197 & 0.371 $\pm$ 0.167 & 0.263 $\pm$ 0.148 & 0.184 $\pm$ 0.130 \\
nh\_white & 0.654 $\pm$ 0.180 & 0.384 $\pm$ 0.173 & 0.269 $\pm$ 0.152 & 0.188 $\pm$ 0.129 \\
other & 0.627 $\pm$ 0.221 & 0.379 $\pm$ 0.186 & 0.271 $\pm$ 0.163 & 0.192 $\pm$ 0.136 \\
\midrule
\multicolumn{5}{c}{\textbf{4 days apart}} \\
\midrule
Group & $k=25$ & $k=50$ & $k=75$ & $k=100$ \\
\midrule
F & 0.707 $\pm$ 0.207 & 0.388 $\pm$ 0.202 & 0.314 $\pm$ 0.175 & 0.213 $\pm$ 0.137 \\
M & 0.659 $\pm$ 0.227 & 0.412 $\pm$ 0.184 & 0.304 $\pm$ 0.150 & 0.226 $\pm$ 0.133 \\
nh\_white & 0.667 $\pm$ 0.227 & 0.414 $\pm$ 0.179 & 0.299 $\pm$ 0.151 & 0.221 $\pm$ 0.129 \\
other & 0.692 $\pm$ 0.204 & 0.401 $\pm$ 0.190 & 0.337 $\pm$ 0.155 & 0.233 $\pm$ 0.144 \\
\bottomrule
\end{tabular}
\caption{Mean churn rates for each group, averaged across all queries and all rankings separated by the same number of days (e.g., one day apart, two days apart, etc.) within $Q_3 \cup Q'_3$.}
\label{tab:q3_churn_cummary}
\end{table}

\begin{table}[ht]
\centering
\begin{tabular}{lcccc}
\toprule
\multicolumn{5}{c}{\textbf{1 day apart}} \\
\midrule
Group & $k=25$ & $k=50$ & $k=75$ & $k=100$ \\
\midrule
F & 0.888 $\pm$ 0.146 & 0.772 $\pm$ 0.153 & 0.667 $\pm$ 0.166 & 0.566 $\pm$ 0.186 \\
M & 0.897 $\pm$ 0.083 & 0.782 $\pm$ 0.114 & 0.674 $\pm$ 0.136 & 0.562 $\pm$ 0.175 \\
nh\_white & 0.890 $\pm$ 0.094 & 0.779 $\pm$ 0.118 & 0.674 $\pm$ 0.137 & 0.567 $\pm$ 0.176 \\
other & 0.893 $\pm$ 0.113 & 0.784 $\pm$ 0.126 & 0.674 $\pm$ 0.156 & 0.562 $\pm$ 0.184 \\
\midrule
\multicolumn{5}{c}{\textbf{2 days apart}} \\
\midrule
Group & $k=25$ & $k=50$ & $k=75$ & $k=100$ \\
\midrule
F & 0.894 $\pm$ 0.162 & 0.798 $\pm$ 0.145 & 0.687 $\pm$ 0.171 & 0.579 $\pm$ 0.204 \\
M & 0.887 $\pm$ 0.092 & 0.789 $\pm$ 0.112 & 0.688 $\pm$ 0.142 & 0.583 $\pm$ 0.179 \\
nh\_white & 0.892 $\pm$ 0.089 & 0.790 $\pm$ 0.113 & 0.687 $\pm$ 0.142 & 0.584 $\pm$ 0.177 \\
other & 0.900 $\pm$ 0.111 & 0.806 $\pm$ 0.126 & 0.691 $\pm$ 0.157 & 0.586 $\pm$ 0.185 \\
\midrule
\multicolumn{5}{c}{\textbf{3 days apart}} \\
\midrule
Group & $k=25$ & $k=50$ & $k=75$ & $k=100$ \\
\midrule
F & 0.907 $\pm$ 0.137 & 0.811 $\pm$ 0.142 & 0.711 $\pm$ 0.171 & 0.600 $\pm$ 0.200 \\
M & 0.891 $\pm$ 0.100 & 0.792 $\pm$ 0.117 & 0.703 $\pm$ 0.143 & 0.604 $\pm$ 0.178 \\
nh\_white & 0.903 $\pm$ 0.086 & 0.801 $\pm$ 0.109 & 0.707 $\pm$ 0.142 & 0.606 $\pm$ 0.177 \\
other & 0.896 $\pm$ 0.106 & 0.792 $\pm$ 0.128 & 0.692 $\pm$ 0.146 & 0.602 $\pm$ 0.179 \\
\midrule
\multicolumn{5}{c}{\textbf{4 days apart}} \\
\midrule
Group & $k=25$ & $k=50$ & $k=75$ & $k=100$ \\
\midrule
F & 0.892 $\pm$ 0.150 & 0.790 $\pm$ 0.135 & 0.726 $\pm$ 0.160 & 0.636 $\pm$ 0.190 \\
M & 0.895 $\pm$ 0.101 & 0.799 $\pm$ 0.116 & 0.704 $\pm$ 0.136 & 0.617 $\pm$ 0.173 \\
nh\_white & 0.899 $\pm$ 0.095 & 0.809 $\pm$ 0.117 & 0.716 $\pm$ 0.138 & 0.622 $\pm$ 0.178 \\
other & 0.913 $\pm$ 0.096 & 0.812 $\pm$ 0.115 & 0.727 $\pm$ 0.149 & 0.636 $\pm$ 0.171 \\
\bottomrule
\end{tabular}
\caption{Mean churn rates for each group for a random ranking, averaged across all queries and separated by the same number of days (e.g., one day apart, two days apart, etc.) within $Q_2 \cup Q_3 \cup Q'_3$.}
\label{tab:random_churn_summary}
\end{table}

\section{Min-Skew Example} \label{app:minskew}
To understand the $\text{MinSkew}@k$ metric better, consider a binary gender setting with the following target proportions: 40\% female and 60\% male. Suppose that at the top-$100$ ranks of a result list, the observed composition is 30 females and 70 males. Then, the observed proportions are:
\[
p^{\tau_r}_{100,\text{female}} = \frac{30}{100} = 0.30, \quad p^{\tau_r}_{100,\text{male}} = \frac{70}{100} = 0.70.
\]
The corresponding skew values for each group are:
\[
\text{Skew}_{\text{female}}@100 = \log\left(\frac{0.30}{0.40}\right) = \log(0.75) \approx -0.2877,
\]
\[
\text{Skew}_{\text{male}}@100 = \log\left(\frac{0.70}{0.60}\right) \approx \log(1.167) \approx 0.154.
\]
Hence, the $\text{MinSkew}@100$ is:
\[
\text{MinSkew}@100 = \min(-0.2877,\ 0.154) = -0.2877.
\]

Now, suppose that after applying a corrective intervention, the top-100 ranks include 39 females and 61 males. Then, $p^{\tau_r}_{100,\text{female}} = 0.39$, and
\[
\text{Skew}_{\text{female}}@100 = \log\left(\frac{0.39}{0.40}\right) = \log(0.975) \approx -0.0253.
\]
In this case, $\text{MinSkew}@100$ improves dramatically to $-0.0253$, showing how even a small absolute change, such as replacing nine male candidates with female candidates, can yield a significant improvement in skew due to the logarithmic nature of the metric.

\end{document}